\documentclass{aa}  

\usepackage{graphicx}
\usepackage{txfonts}
\usepackage{siunitx}
\DeclareSIUnit\gauss{G}
\usepackage{ragged2e}
\usepackage{natbib}
\usepackage{subfigure}
\usepackage{hyperref}
\hypersetup{
    colorlinks=true,
    linkcolor=blue,
    citecolor=blue,
    filecolor=magenta,      
    urlcolor=cyan,
}

\newcommand{\vae}{v_{\mathrm{A,e}}}
\newcommand{\vai}{v_{\mathrm{A,i}}}

\newcommand{\vk}{v_{\rm k}}

\newcommand{\rhoi}{\rho_{\rm i}}
\newcommand{\rhoe}{\rho_{\rm e}}

\newcommand{\rhotr}{\rho_{\rm tr}}

\begin{document}

   \title{Nonlinear  evolution of  fluting oscillations in  coronal flux tubes}

   \author{Roberto Soler\inst{1,2} \and Andrew Hillier\inst{3}
          }

   \institute{Departament de F\'isica, Universitat de les Illes Balears, E-07122, Palma de Mallorca, Spain\\ \and Institut d'Aplicacions Computacionals de Codi Comunitari (IAC3), Universitat de les Illes Balears, E-07122, Palma de Mallorca, Spain \\ \and Department of Mathematics \& Statistics, University of Exeter, Exeter EX4 4QF, UK}
   
   


  \abstract{Magnetic flux tubes in the solar corona support a rich variety of transverse oscillations, which are theoretically interpreted as magnetohydrodynamic (MHD)  modes with a fast and/or Alfv\'enic character. In the standard flux tube model made of a  straight cylindrical  tube, these modes can be classified according to their azimuthal wavenumber, $m$. Sausage $m=0$ modes produce periodic expansion and contraction of the tube cross section and are observed during solar flares. Kink $m=1$ modes laterally displace the tube axis and are related to, for example, post-flare global transverse oscillations of coronal loops. Fluting $m \geq 2$ modes produce  disturbances that are mainly confined to the tube boundary, but their observation remains elusive to date. We use 3D ideal MHD numerical simulations to investigate the nonlinear evolution of fluting modes in  coronal flux tubes with transversely nonuniform boundaries. The simulations show that fluting modes are short-lived as coherent, collective motions of the  flux tube. Owing to the process of resonant absorption, fluting oscillations become overdamped modes  in tubes with wide enough nonuniform boundaries. During the nonlinear evolution, shear flows drive  the Kelvin-Helmholtz instability at the tube boundary, which further disrupts the coherent fluting oscillation. For large-enough oscillation amplitudes, baroclinic instabilities of Rayleigh-Taylor type are also present at locations in the boundary where the plasma acceleration is normal to the boundary. The evolution of the instabilities drives turbulence in the flux tube, which may inhibit the resonant damping. However, the oscillations remain strongly damped even in this case. As a result of the combination of the  strong damping and the induced instabilities, it is unlikely that coronal flux tubes can support fluting modes as sufficiently enduring coherent oscillations.\\
    {\bf Note on the preprint version: Movies are accessible here:} \url{https://owncloud.uib.es/index.php/s/zSrdXwkJAyfGwkD}}


   \maketitle

\section{Introduction}

Oscillations in magnetic flux tubes of the solar atmosphere are usually interpreted as magnetohydrodynamic (MHD)  modes \citep[see, e.g.,][]{roberts2019}. In the standard  model made of a straight magnetic cylinder with a uniform axial magnetic field, the linear MHD modes can be  classified as transverse or longitudinal modes, according to the polarization of their velocity field with respect to the flux tube axis. In addition, standing modes can also be classified according to the number of nodes that their eigenfunctions have in the radial and longitudinal directions. In this paper, we focus on radially and longitudinally fundamental transverse modes and the following discussion applies to those modes alone.

The modes display very different properties depending on the value of the azimuthal wavenumber, $m$. Thus,  modes with $m=0$ are called torsional modes and sausage modes. Torsional modes are pure Alfv\'en waves polarized in the azimuthal direction and localized on specific magnetic surfaces that are decoupled from each other. In transversely nonuniform tubes, torsional Alfv\'en modes undergo phase mixing and do not produce coherent oscillations of the flux tube \citep[see, e.g.,][]{Heyvaerts1983,diaz2021}.  Conversely, sausage waves are global fast magnetoacoustic modes that cause the periodic expansion and contraction of the loop cross section. They are observed during solar flares as quasi-periodic oscillations \citep[see the review by][]{zimovets2021}. In straight magnetic tubes, torsonal Alfv\'en modes and fast sausage modes are decoupled from each other, although they have the same  azimuthal symmetry. A magnetic twist, however, couples both modes \citep[see, e.g.,][]{goossens2011}.

The global modes with $m=1$ are called kink modes. These modes produce the transverse displacement of the tube axis and have been linked to the observed transverse oscillations of coronal loops \citep[e.g.,][]{nakariakov1999,aschwanden1999}, prominence threads \citep[e.g.,][]{okamoto2007,lin2009}, and chromospheric waveguides \citep[e.g.,][]{McIntosh2011,jess2012}, among other examples. Kink modes in transversely nonuniform tubes have mixed fast and Alfv\'en properties, but in thin tubes (TTs) the Alfv\'en part dominates. Furthermore, kink modes are resonant in the Alfv\'en continuum. For these reasons, kink modes have been called Alfv\'enic \citep[see][]{goossens2009,goossens2012}, a term that was first introduced by \citet{ionson1978}. An excellent review on Alfv\'enic waves in the solar atmosphere is given in \citet{morton2023}.   Kink modes undergo resonant damping and feed their energy to the $m=1$ Alfv\'en continuum modes that, in turn, undergo phase mixing \citep[see, e.g.,][]{terradas2006,solerterradas2015}. The nonlinear evolution of this process is characterized by the  triggering of the Kelvin-Helmholtz instability (KHi) when the generated azimuthal shear flows  are strong enough \citep[see, e.g.,][]{terradas2008, soler2010}. The temporal evolution obtained from numerical simulations shows the subsequent transition of the flux tube into a turbulent state in which the mixing of the internal and external plasmas occurs.

The modes with $m \geq 2$ are called fluting modes.  These modes do not displace the tube axis, but produce perturbations that are mainly localized around the tube boundary in a way that is reminiscent of surface waves. As in the case of kink waves, fluting modes are resonant in the Alfv\'en continuum and have a predominantly   Alfv\'enic character. In view of these properties, their evolution should share most of the complex dynamics of kink waves explained above.  In spite of this,  not many wave studies in the solar corona have included results on fluting modes \citep[see a few examples in,  e.g.,][]{erdelyi2010,morton2011,ruderman2017,Shukhobodskaia2021}. To our knowledge,  the  only dedicated study of resonantly damped fluting modes is presented in \cite{soler2017} using linear theory.   However, an investigation of the nonlinear  evolution of fluting modes is lacking in the literature. As a matter of fact, the theoretical study of fluting modes is  lagging behind those of sausage and kink modes for one fundamental reason: there are no observations of fluting modes in the solar corona. So, within the rich family of transverse MHD modes of a coronal magnetic flux tube, fluting modes appear to be the forgotten members. 

The goal of this paper is to fill this gap in the literature and investigate the temporal evolution of fluting modes in a transversely nonuniform coronal flux tube model. This work is a natural follow-up of \citet{soler2017}, in which fluting modes were studied with an analytic method in linear MHD. Here, we perform nonlinear MHD numerical simulations.   Another aim of the paper is to offer a potential explanation for why the detection of fluting modes has been so elusive. One likely reason is that the perturbations linked to fluting modes are below the spatial resolution capability of current instruments. However, even if future instruments achieve the required high resolution, we argue throughout the paper that the specific dynamics of fluting modes make them unlikely to be observed as sustained, coherent oscillations of the flux tube.

\section{Model and numerical method}

We used the so-called standard model to represent a coronal loop. The model is composed of a straight magnetic cylinder of radius $R$ and length $L$. We considered $L/R= 20$. The reason for considering a shorter tube  than the typically observed loop lengths is to speed up the simulation times, since the periods of the oscillations are proportional to the  length. However, regarding the overall dynamics of the simulations, we do not expect significant changes if a longer loop was considered. For torsional oscillations, the effect of the loop length is explored in \citet{diaz2021}.   The tube was embedded in a uniform low-$\beta$ plasma, where $\beta$ refers to the ratio of the gas pressure to the magnetic pressure.  The gas pressure, $p$, was uniform. For convenience, we used cylindrical coordinates to define the model, so that $r$, $\varphi$, and $z$ denote the radial, azimuthal, and longitudinal coordinates, respectively. A straight magnetic field was assumed along the axis of the cylinder; namely, ${\bf B} = B {\bf 1}_z$, where $B$ is a constant. The ends of the cylinder, located at $z=\pm L/2$, were fixed at two rigid walls representing the solar photosphere. This model neglects the presence of the chromosphere at the loop feet. 

\begin{figure}[!htb]
    \centering
    \includegraphics[width=0.8\columnwidth]{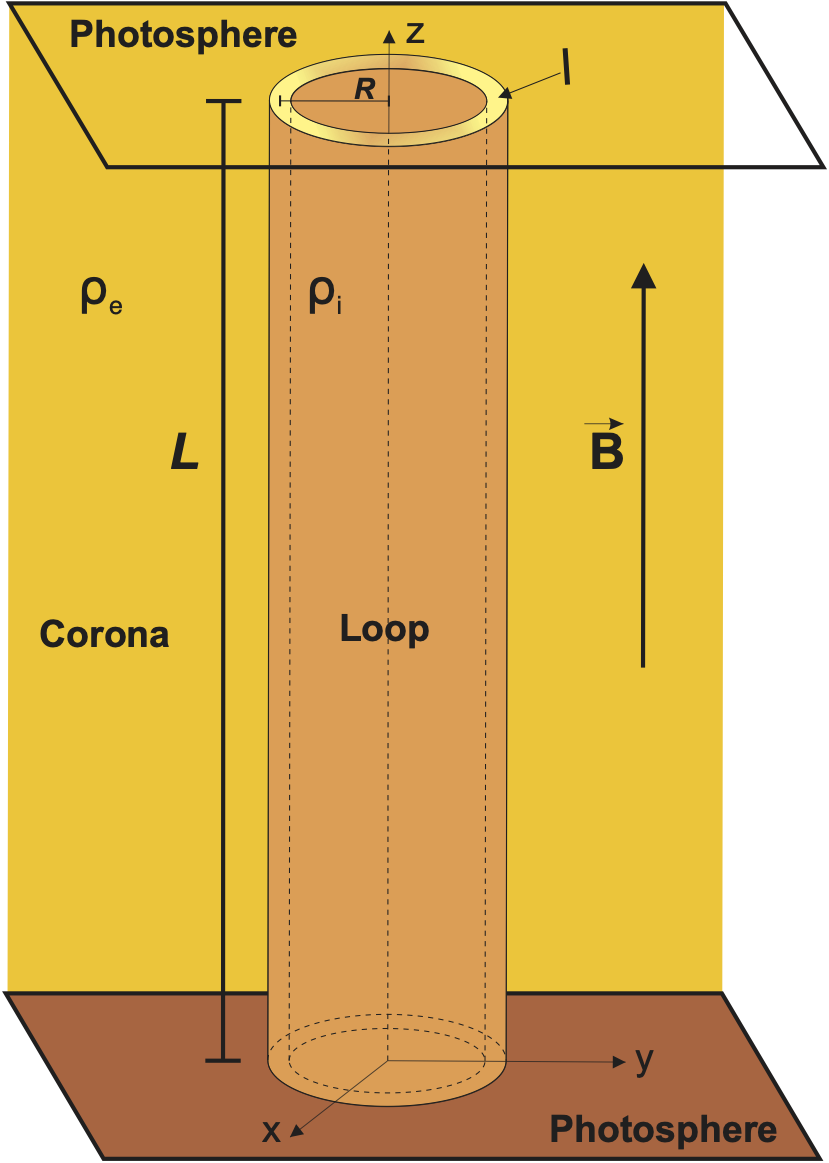} 
    \caption{Sketch of the coronal loop model used in this work.}
    \label{fig:model}
\end{figure}

The effect of gravitational stratification was ignored and the mass density, $\rho = \rho(r)$, was assumed to be nonuniform in the radial direction alone. We used the following radial dependence for the density:
\begin{equation}
 \rho(r) = \left\{
\begin{array}{lll}
\rhoi, & \textrm{if} & r \leq R - l/2, \\
\rhotr(r), & \textrm{if} & R -l/2 < r < R +  l/2,\\
\rhoe, & \textrm{if} & r \geq R+l/2,
\end{array}
\right.
\end{equation}
where $\rhoi$ and $\rhoe$ are internal and external uniform densities and $\rhotr(r)$ is a continuous density profile that connects the internal medium and the external medium. We considered a sinusoidal transition; namely,
\begin{equation}
\rhotr(r) = \frac{\rhoi}{2} \left[\left( 1 + \frac{\rhoe}{\rhoi} \right) -  \left( 1 - \frac{\rhoe}{\rhoi} \right) \sin \left(\frac{\pi}{l}(r-R) \right) \right]. \label{eq:transition}
\end{equation}
We assumed $\rhoi > \rhoe$ to represent a loop that is denser than the background coronal plasma.  The width of the nonuniform transition, $l$, can take any value between $l = 0$ (abrupt jump) and  $l = 2R$  (fully nonuniform tube).  A sketch of the model is displayed in Fig.~\ref{fig:model}.

We used the PLUTO code \citep{mignone2007} to numerically solve the 3D nonlinear ideal MHD equations; namely,
\begin{eqnarray}
\frac{{\rm D}\rho}{{\rm D}t} & = & - \rho \nabla \cdot {\bf v}, \\
\rho\frac{{\rm D}{\bf v}}{{\rm D}t} & = & - \nabla p + \frac{1}{\mu_0} \left( \nabla \times {\bf B} \right) \times {\bf B}, \\
\frac{\partial{\bf B}}{\partial t} &= & \nabla \times \left( {\bf v} \times {\bf B} \right), \\
\frac{{\rm D}p}{{\rm D}t} & = & \frac{\gamma p}{\rho} \frac{{\rm D}\rho}{{\rm D}t},
\end{eqnarray}
where ${\bf v}$ is the velocity, $\gamma$ is the adiabatic index, $\mu_0$ is the magnetic permeability, and $\frac{{\rm D}}{{\rm D}t}  = \frac{\partial}{\partial t}  + {\bf v} \cdot \nabla$ is the Lagrangian derivative. The remaining symbols have been defined before. We used a shock-capturing finite-difference spatial reconstruction based on the fifth-order WENOZ scheme \citep{borges2008} together with the  hlld approximate Riemann solver \citep{miyoshi2005} for flux computation. The temporal evolution was performed with an un-split third-order Runge Kutta scheme. The solenoidal constraint on the magnetic field was enforced with the hyperbolic divergence cleaning method \citep{dedner2002}.   

The code solves the MHD equations in  Cartesian coordinates. The computational domain extends from $-2.5R$ to $2.5R$ in the $x$ and $y$ directions, and from $-L/2$ to $L/2$ in the $z$ direction. We considered outflow boundary conditions -- that is, zero gradients -- at the lateral boundaries to allow emitted waves, if any, to leave the domain. We checked that no noticeable reflections occur at those boundaries  and that the dynamics that develop during the simulations occur sufficiently far from the boundaries. Conversely,  line-tying conditions are imposed at the top and bottom boundaries to represent the anchoring of the magnetic field lines in the photosphere. The computational box has a numerical resolution of 301×301×51 points, which are distributed uniformly.  We ran tests with higher resolutions (not included here) and checked that  this numerical resolution is  enough for our purpose of studying the damping of global modes and allows us to perform a large number of simulations within reasonable running times.

At $t=0$, we aim to excite transverse fluting oscillations of the coronal loop. To this end, we perturbed the system with a velocity field that is  close to that of the  fundamental fluting eigenmode in the case of a loop with no boundary layer; that is,  with $l=0$. The trapped MHD  modes of a flux tube with an abrupt boundary have been studied in detail \citep[see, e.g.,][]{edwin1983,goossens2009,goossens2012,roberts2019}. The radial, $v_r$, and azimuthal, $v_\varphi$,  velocity  perturbations corresponding to the fundamental mode along the loop can be written as
\begin{eqnarray}
v_r (r,\varphi,z) &=& \tilde{v}_r (r) \cos \left( m \varphi \right) \cos\left( \frac{\pi}{L} z \right) , \\
v_\varphi(r,\varphi,z)  &=& - \tilde{v}_\varphi (r) \sin \left( m \varphi \right) \cos\left( \frac{\pi}{L} z \right), 
\end{eqnarray}
where $m$ is the azimuthal wavenumber that has already been defined before and $\tilde{v}_r (r)$ and $\tilde{v}_\varphi (r)$ contain the radial dependence of the radial and azimuthal velocity components, respectively, which generally depend on Bessel functions in the internal plasma and modified Bessel functions in the external plasma \citep[see details in, e.g.,][]{roberts2019}. For a TT -- that is, when $L/R \gg 1$, and for $m\neq 0$ -- the radial dependence is well approximated by
\begin{equation}
\tilde{v}_r (r) = v_0 \left\{ \begin{array}{lll}
                            \left( r/R \right)^{m-1}, & \textrm{if}, & r \leq R, \\
                             \left( R/r \right)^{m+1}, & \textrm{if}, & r > R,
                            \end{array} \right.\label{eq:vr}
\end{equation}
\begin{equation}
\tilde{v}_\varphi (r) = v_0 \left\{ \begin{array}{lll}
                            \left( r/R \right)^{m-1}, & \textrm{if}, & r < R, \\
                            - \left( R/r \right)^{m+1}, & \textrm{if}, & r > R,
                            \end{array} \right.\label{eq:vfi1}
\end{equation}
where $v_0$ is an arbitrary  amplitude. In addition,  we imposed for simplicity that the longitudinal velocity perturbation vanishes initially; namely,
\begin{equation}
   v_z(r,\varphi,z)  = 0,
\end{equation}
which is a reasonable approximation for transverse waves in the low-$\beta$ regime applicable in the solar corona. Longitudinal velocities are later generated during the evolution of the oscillations, although their magnitude remains much smaller than the transverse velocities. Some plots of the radial and azimuthal components of the Lagrangian displacement, which are proportional to the respective velocity components, can be seen in \citet{soler2017} in the case of fluting modes.

We used the above analytic prescription for the velocity  as the initial condition in the code even when $l\neq 0$. However, we note that according to Eqs.~(\ref{eq:vr}) and (\ref{eq:vfi1}),  $\tilde{v}_r (r)$ is continuous but  $\tilde{v}_\varphi (r)$  is discontinuous at $r=R$, which may pose a problem from a numerical point of view. Therefore, in our numerical implementation we replaced Eq.~(\ref{eq:vfi1}) with
\begin{equation}
\tilde{v}_\varphi (r) =v_0 \left\{ \begin{array}{lll}
                            \left( r/R \right)^{m-1}, & \textrm{if}, & r < R-l/2, \\
                            f(r), & \textrm{if}, & R-l/2 \leq r \leq R+l/2, \\
                             - \left( R/r \right)^{m+1}, & \textrm{if}, & r > R + l⁄2,
                            \end{array} \right.
\end{equation}
with
\begin{equation}
    f(r) = \left( 1- \frac{l}{2R} \right)^{m-1} - \left[\left( 1 + \frac{l}{2R} \right)^{-m+1} + \left( 1- \frac{l}{2R} \right)^{m-1}\right] \frac{r-R+l/2}{l},
\end{equation}
so that the jump of $\tilde{v}_\varphi (r)$ at $r=R$ is avoided by connecting the internal and external profiles with a linear ramp in the nonuniform transitional layer.

\section{Previous theoretical considerations}
\label{sec:prevtheo}

Before discussing the numerical simulations,  we shall enumerate some considerations that can help us understand the simulation dynamics and put their results in the appropriate context. First, we shall revisit  the computations of the fluting quasi-modes done in \citet{soler2017}. When $l\neq 0$, transverse MHD modes with $m\neq 0$ are resonant in the Alfv\'en continuum and have complex frequencies because of resonant damping; namely, $\omega = \omega_{\rm R} + i \omega_{\rm I}$. These resonantly damped modes are not true eigenmodes of ideal MHD, hence the name quasi-mode or virtual mode \citep[see, e.g.,][]{sedlacek1971,rae1981,goedbloed1983,poedts1991,soler2022}. They physically represent a transient collective oscillation of the plasma and are the descendants of the undamped, true eigenmodes found when $l=0$. The lifetime of the collective motion is associated with  the quasi-mode damping rate, $\left|\omega_{\rm I}\right|$.  The process  that leads to the quasi-mode damping is of a linear nature. Because of resonant absorption, the energy of the collective oscillation  is transferred to the nonuniform boundary of the tube in the form of local Alfv\'en waves with the same azimuthal symmetry as that of the original collective motion. After a timescale given by the inverse of the quasi-mode damping rate, $\sim 1/\left|\omega_{\rm I}\right|$,   the flux tube motion is no longer global. Plasma motions become uncoordinated and localized at the nonuniform boundary, where the concurrent process of phase mixing cascades the wave energy toward smaller and smaller scales \citep[see, e.g,][]{Heyvaerts1983,Browning1984,Cally1991,solerterradas2015,diaz2021}.  In the context of  solar atmospheric flux tubes, the damping of global transverse oscillations has theoretically been studied in detail in the case of kink ($m=1$) modes \citep[see, e.g.,][]{goossens1992,ruderman2002,vandoorsselaere2004,andries2005,soler2013,solergoossens2024}, but the same mechanism  equally applies to all fluting ($m\geq 2$) modes. Indeed, as is shown in \citet{soler2017}, the process turns out to be more efficient, and so faster, in the case of fluting modes.

\begin{figure}[!tb]
    \centering
    \includegraphics[width=0.9\columnwidth]{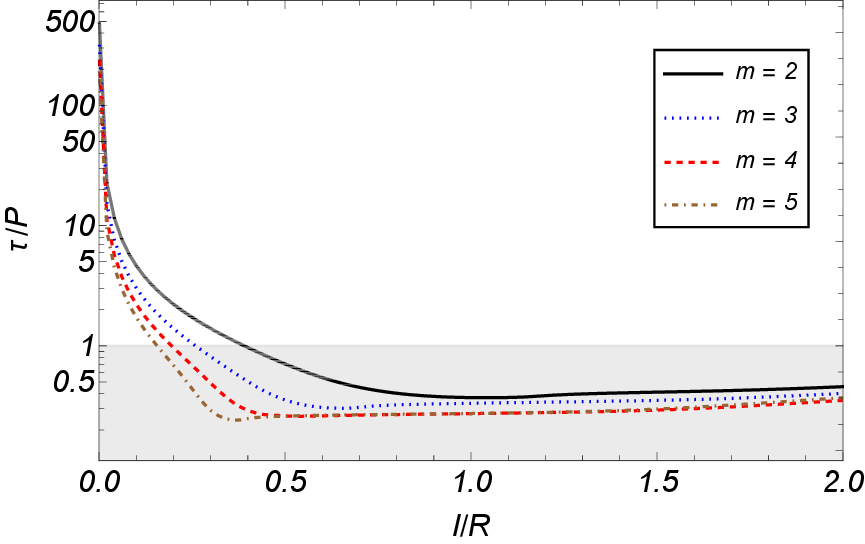} 
    \caption{Ratio of the exponential damping time, $\tau$, to the oscillation period, $P$, as a function of $l/R$ for the fluting quasi-modes with $m=$~2, 3, 4, and 5. The gray area corresponds to $\tau / P < 1$, i.e., overdamped quasi-modes. Results were computed with the Frobenius method explained in \citet{soler2017}, considering $L/R = 20$ and $\rhoi/\rhoe = 5$.}
    \label{fig:frobenius}
\end{figure}

Figure~\ref{fig:frobenius} displays the ratio of the amplitude exponential damping time, $\tau = 1/\left| \omega_{\rm I} \right|$, to the oscillation period, $P = 2\pi / \omega_{\rm R}$, as a function of $l/R$ for the fundamental fluting quasi-modes with $m=$~2, 3, 4, and 5. These results were obtained with the Frobenius method described generally in \citet{soler2013} and applied to fluting modes in \citet{soler2017}, where the interested readers can find all the details. For these computations, we used $L/R = 20$ and $\rhoi/\rhoe = 5$. The shaded area in  Fig.~\ref{fig:frobenius} corresponds to $\tau /P < 1$, so that the quasi-modes are overdamped. Physically, this implies that most of the energy of the global motion has already  flown to the nonuniform layer before a single period of the collective oscillation has been completed. As a consequence of that, the coherent phase of the flux tube motion  is so short-lived that it can hardly be called an ``oscillation'' in the usual sense of an enduring periodic motion.  

For $m\neq 0$, expressions for $P$ and $\tau /P$ in the TT ($L/R \gg 1$)  and thin boundary (TB, $l/R \ll 1$) approximation are \citep[see][]{soler2017}
\begin{eqnarray}
P &=& \frac{2 L}{\vk} \equiv P_k, \label{eq:per} \\
    \frac{\tau}{P} &=& \frac{2}{\pi} \frac{R}{l} \frac{\rhoi+\rhoe}{\rhoi-\rhoe} \frac{1}{m}, \label{eq:taup}
\end{eqnarray}
with
\begin{equation}
    \vk = \sqrt{\frac{\rhoi\vai^2 + \rhoe\vae^2}{\rhoi + \rhoe}},
\end{equation}
where $\vai = B/\sqrt{\mu_0 \rhoi}$ and $\vae = B/\sqrt{\mu_0 \rhoe}$ are the internal and external Alfv\'en velocities, $\mu_0$ is the vacuum magnetic permeability, and the factor $2/\pi$ in Eq.~(\ref{eq:taup}) appears because of the use of a sinusoidal profile in the nonuniform layer. This factor would be different if another profile were used \citep[see][]{soler2014}. According to Eq.~(\ref{eq:per}), the period of all quasi-modes with $m\neq 0$ is the same in the TT and TB approximation. However, Eq.~(\ref{eq:taup}) shows that quasi-modes with different $m$ have different damping times. Equation~(\ref{eq:taup}) is most frequently given in the literature in the case of kink waves ($m=1$) and usually the factor $1/m$ is not explicitly written. The $1/m$ dependence plainly evidences that resonant damping is more efficient for fluting modes than for kink modes and gets more and more efficient as $m$ increases. 

We note that Eq.~(\ref{eq:taup}) is theoretically not applicable beyond the TB condition, although its accuracy may still be sufficiently good \citep[see][]{vandoorsselaere2004,soler2014}. Conversely, the results of  Fig.~\ref{fig:frobenius} are obtained with a more general approach that is not restricted by the TB approximation. Figure~\ref{fig:frobenius} shows that the quasi-modes get into the overdamped region for relatively small values of $l/R$ for which Eq.~(\ref{eq:taup}) is  still applicable. Hence, Eq.~(\ref{eq:taup}) can be used to easily estimate the critical values of $l/R$ for which the quasi-modes become overdamped; namely,
\begin{equation}
    \frac{l}{R} \approx \frac{2}{\pi} \frac{\rhoi+\rhoe}{\rhoi-\rhoe} \frac{1}{m}. \label{eq:critlR}
\end{equation}
For $\rhoi/\rhoe = 5$, Eq.~(\ref{eq:critlR}) gives $l/R \approx$~0.48, 0.32, 0.24, and 0.19 for $m=$~2, 3, 4, and 5, respectively, which is in good agreement with the numerical results of Fig.~\ref{fig:frobenius}. 

The presence of the factor $\frac{\rhoi+\rhoe}{\rhoi-\rhoe}$ in Eq.~(\ref{eq:critlR})  indicates that the density ratio, $\rhoi/\rhoe$, may play a role regarding the critical value of $l/R$. When $\rhoi/\rhoe \to 1$, we have $\frac{\rhoi+\rhoe}{\rhoi-\rhoe} \gg 1$, which suggests that quasi-modes in loops with  low-density ratios may require wider nonuniform transitions  than in denser loops to become overdamped.  In other words: fluting modes are less prone to being heavily damped in loops with low-density contrast.

From the above analysis,  we can make some predictions about the temporal evolution. We can anticipate that coherent fluting oscillations in nonuniform loops should be very short-lived, unless the nonuniform boundary layer is very thin and/or the density ratio is remarkably low. In addition,  the larger the azimuthal wavenumber, $m$, the briefer the coherent motion would be. However, nonlinear effects, which are not included in the above quasi-mode linear analysis, can also heavily affect the evolution of fluting oscillations. It is well known from  numerical simulations of kink oscillations \citep[e.g.,][]{terradas2008,antolin2014,howson2017,hillier2019} and torsional Alfv\'en oscillations \citep[e.g.,][]{guo2019,diaz2021,diaz2022} that the Kelvin-Helmholtz instability (KHi) is eventually triggered at the loop boundary either directly by the kink mode perturbations or when the phase-mixing shear flows are strong enough \citep[see, e.g.,][]{Browning1984,soler2010}. The nonlinear evolution of the KHi drives plasma mixing and turbulence in the flux tube. It is likely that the KHi is also triggered during the evolution of fluting modes and its nonlinear development at the loop boundary may potentially have a  strong impact on the fluting oscillations themselves. It is precisely at the loop boundary where the perturbations associated with fluting modes are mainly localized \citep[see Fig.~2 of][]{soler2017}.

Therefore, we argue that not only the resonant damping rate may be important in setting the timescale for the existence of a coherent fluting oscillation, but the KHi  should also be accounted for. Which one of the two effects dominates would likely depend upon the loop properties and the amplitude of the initial perturbation. For instance, in loops with thin nonuniform transitions the quasi-mode damping time would be large, but the steep boundary of the loop would be prone to quickly become KH-unstable by a direct instability. In this scenario, we speculate that the flux tube boundary may be  disrupted by the KHi and its associated turbulence before resonant absorption has completely drained the energy from the quasi-mode. This may cause the resonant damping to be less efficient than what the linear analysis predicts  \citep{hillier2023}. Conversely, in a loop with a thick nonuniform layer, a direct instability is less likely to happen, but the KHi can still be driven by the phase mixing flows. In this case, the  damping of the quasi-mode would be fast, as linearly predicted, and the KHi onset would be delayed to a later time because of the slow development of phase mixing within the nonuniform boundary. Thus, in a loop with a thick boundary the coherent fluting motion would simply attenuate because of resonant absorption, whereas the KHi and turbulence could even make their entrance well after the global fluting oscillation has decayed. 

In addition to the above considerations, the amplitude of the initial velocity perturbation should also influence how fast the KHi sets in. We expect that the larger the initial amplitude, the quicker the KHi would be able to grow. The quasi-mode damping may be affected by the development of the turbulent region,  potentially inhibiting the resonant damping or decreasing its efficiency.  The numerical simulations should confirm or correct these anticipated results based on  theoretical arguments.

\section{Weakly nonlinear oscillations}
\label{sec:lowamp}


\begin{figure}[!tb]
    \centering
    \includegraphics[width=0.99\columnwidth]{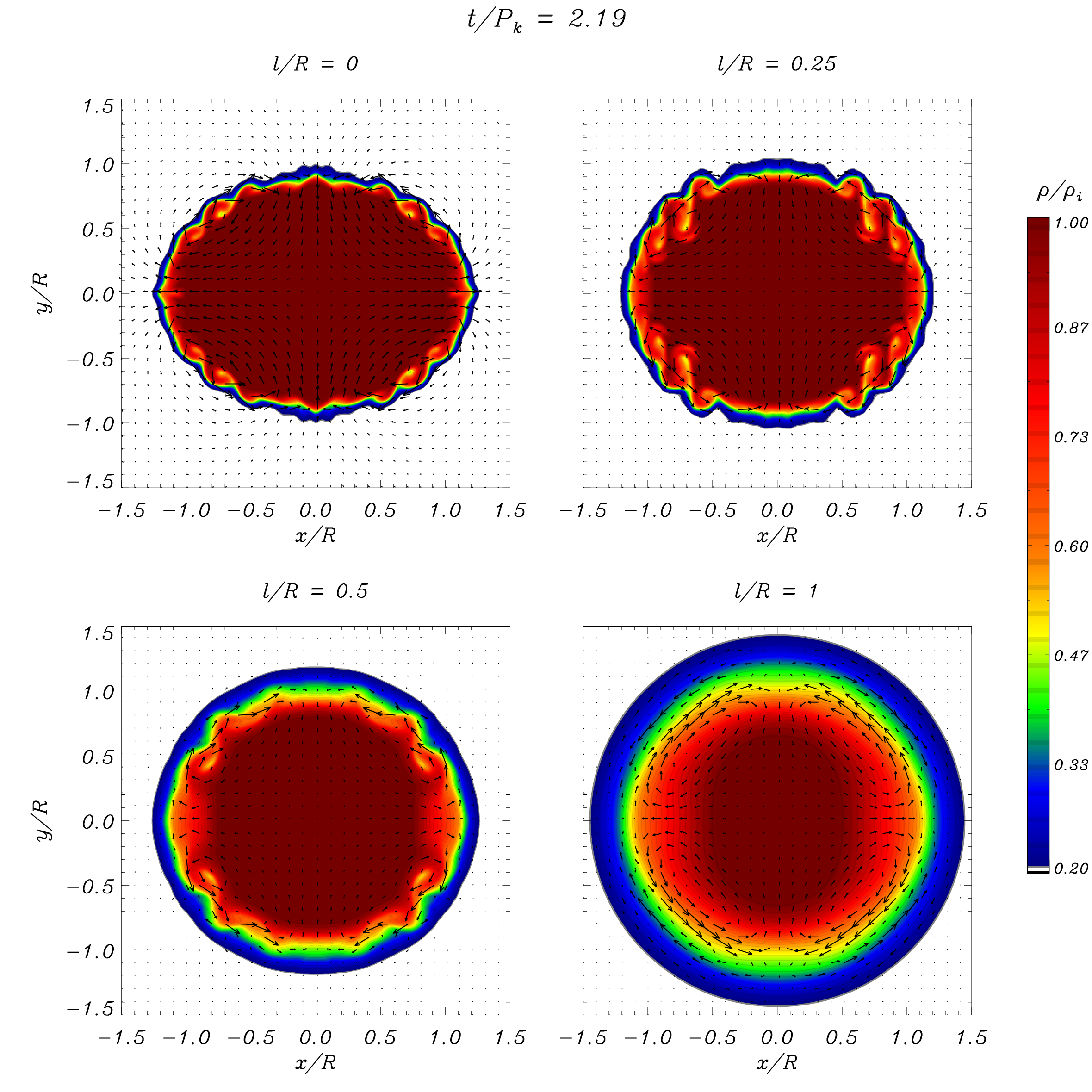} 
    \caption{Simulations of the $m=2$ mode with  $v_0/\vai = 0.05$ and  $\rhoi/\rhoe = 5$. Cross-sectional cuts of the density at the tube center, $z=0$, for the cases with $l/R=0$ (top left), $l/R=0.25$ (top right), $l/R=0.5$ (bottom left), and $l/R=1$ (bottom right). Only a subregion of the complete numerical domain in the vicinity of the flux tube is shown. A  snapshot of the evolution at $t /P_k=2.19$ is displayed in the still image. The complete temporal evolution is available in the accompanying movie.}
    \label{fig:simulm2varl}
\end{figure}

Here, we present and discuss the results of the numerical simulations. Length, velocity, and time have been normalized with respect to $R$, $\vai$, and $P_k$, respectively, where $P_k$ denotes the period  in the TT approximation given in Eq.~(\ref{eq:per}). Density was normalized with respect to the internal density, $\rhoi$. The simulations have been run up to a maximum time corresponding to $5 P_k$. 

Firstly, we considered the case in which the oscillations are weakly nonlinear, so that a comparison with the linear results presented in the previous section could be made. To this end, we used an initial velocity amplitude of $v_0/\vai = 0.05$. Larger values of the amplitude were considered later to account for more pronounced nonlinear effects.

\subsection{Effect of the nonuniform layer width and the density contrast}

\begin{figure*}[!tb]
    \centering
    \includegraphics[width=1.95\columnwidth]{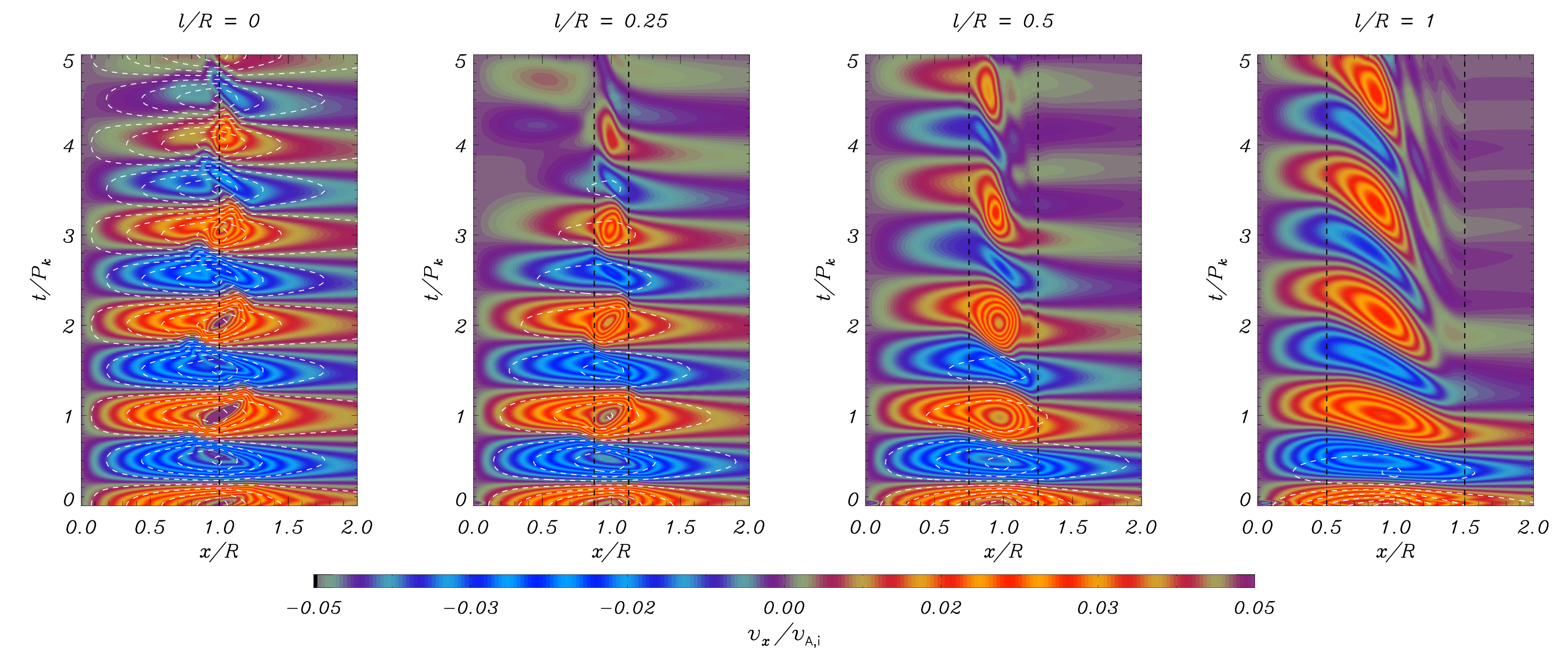} 
    \caption{Time-distance diagrams of the $x$ component of velocity at $y=z=0$ and $x\in[0,2R]$ for the simulations of the $m=2$ mode with $v_0/\vai = 0.05$ and $\rhoi/\rhoe = 5$ . From left to right, the four panels correspond to the results with $l/R=$~0, 0.25, 0.5, and 1, respectively. The vertical dashed black lines denote the boundaries of the nonuniform region. The dashed white contours represent the expected behavior of the quasi-mode in each case.}
    \label{fig:timedistancem2varl}
\end{figure*}

\begin{figure*}[!tb]
    \centering
    \includegraphics[width=1.95\columnwidth]{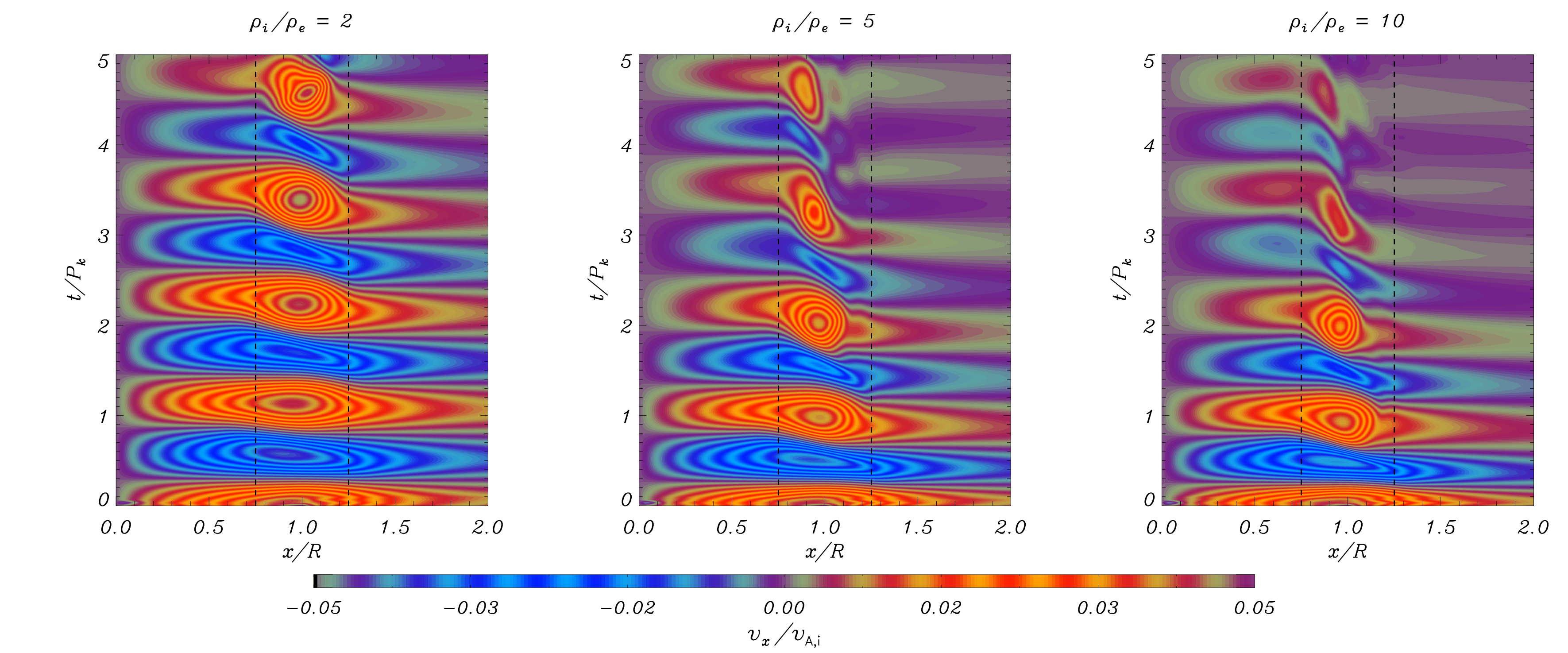} 
    \caption{Time-distance diagrams of the $x$ component of velocity at $y=z=0$ and $x\in[0,2R]$ for the simulations of the $m=2$ mode with  $v_0/\vai = 0.05$ and $l/R = 0.5$. From left to right, the three panels correspond to the results with $\rhoi/\rhoe =$~2, 5, and 10, respectively. The vertical dashed black lines denote the boundaries of the nonuniform region. }
    \label{fig:timedistancem2varrho}
\end{figure*}

To start with, we considered the mode with $m=2$ and used a density contrast of $\rhoi/\rhoe = 5$. We performed a series of four simulations in which we progressively increased the thickness of the nonuniform layer of the tube. We considered $l/R =$~0, 0.25, 0.5, and 1. In order to compare the results, Fig.~\ref{fig:simulm2varl} displays cross-sectional cuts of the density at the tube center, $z=0$, for the four simulations. The transverse velocity field is also represented by means of the superimposed  arrows.  Figure~\ref{fig:simulm2varl}  shows a particular snapshot of the evolution taken at $t =2.19 P_k$, whereas the complete temporal evolution can be seen in the accompanying animation.

Initially, the behavior is similar for the four considered values of $l/R$. The cross-sectional area of the tube suffers periodic deformations consistent with an oscillation with $m=2$ azimuthal symmetry. However,   differences  appear between the simulations as time increases. The simulation with $l/R=0$ requires a separate discussion, so we put the case with $l/R=0$ aside for the moment. In the remaining cases, we find that the oscillations are damped in time and the damping gets stronger when $l/R$ increases. This is consistent with the linear theory of resonant damping (Fig.~\ref{fig:frobenius}). As resonant absorption transfers wave energy to the nonuniform boundary of the tube, azimuthal shear flows are generated there due to the concurrent process of phase mixing. The presence of shear flows in that region is clearly evidenced by the velocity field. For the same simulation time, the larger $l/R$, the smaller the amplitude of the phase-mixing shear flows. The KHi is eventually triggered in the simulations, with the onset time depending on the value of $l/R$. The onset of the KHi is analyzed in more detail later.

In the case with $l/R=0$,  there is an abrupt density jump at the boundary of the tube initially. However, the dynamics of the simulation quickly generates a very thin nonuniform transition. This was also noted in the kink mode simulations by \citet{antolin2019}. The finite resolution of the numerical mesh and its associated numerical diffusion also play a role in this process. The oscillations in this case are characterized by the presence of very weak damping. Although a thin transitional layer is dynamically generated, the resolution of the numerical mesh might not be enough to correctly describe the resonant absorption that should occur in this thin nonuniform transition. As such, the weak damping is most likely caused by the induced KHi, which extracts energy from the global oscillation. In the case of kink modes, this has been explored by \cite{VanDoorsselaere2021}.

 Unlike kink modes, fluting modes displace neither the axis of the flux tube nor the center of mass. This causes the investigation of the collective oscillation damping to be more laborious in the case of fluting modes. To study how the transient collective  oscillation evolves, we consider the  cross-sectional cut of Fig.~\ref{fig:simulm2varl} and, at every time step, sample the $x$ component of velocity along $y=0$ and for $x\in[0,2R]$. We consider this particular direction for the sampling because no KHi vortices form initially along the $y=0$ direction, and so we can partially remove the effect of the KHi from the analysis. We note that the $x$ component of velocity along the considered line is indeed the radial component of velocity in cylindrical coordinates.  With these data, we plot in Fig.~\ref{fig:timedistancem2varl} time-distance diagrams for the four considered values of $l/R$. Positive (negative) values of velocity are represented by orange (blue) colors. To compare with the linear results, in each panel we overplot with dashed white contours the expected behavior of the quasi-mode for that particular value of $l/R$. The quasi-mode temporal dependence is proportional to $\cos\left( 2\pi t/P \right)\exp\left( -  t ⁄\tau \right)$, where $P$ and $\tau$ are the quasi-mode period and damping time that are computed with the same method as that used for the results of Fig.~\ref{fig:frobenius}.   The quasi-mode analysis predicts an undamped  oscillation ($\tau\to\infty$) for $l/R=0$, an underdamped oscillation ($\tau> P$) for $l/R=0.25$, and an overdamped mode ($\tau < P$) for $l/R=1$. In turn, the case with $l/R=0.5$ approximately corresponds to the critically damped scenario ($\tau \approx P$) separating the underdamped and overdamped regimes.

\begin{figure*}[!tb]
    \centering
     \includegraphics[width=0.9\columnwidth]{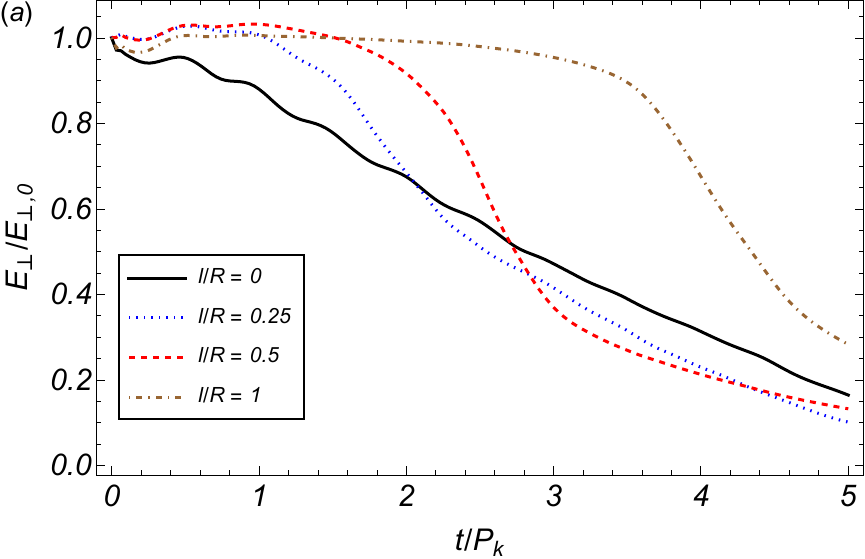} \hspace*{2ex}
      \includegraphics[width=0.9\columnwidth]{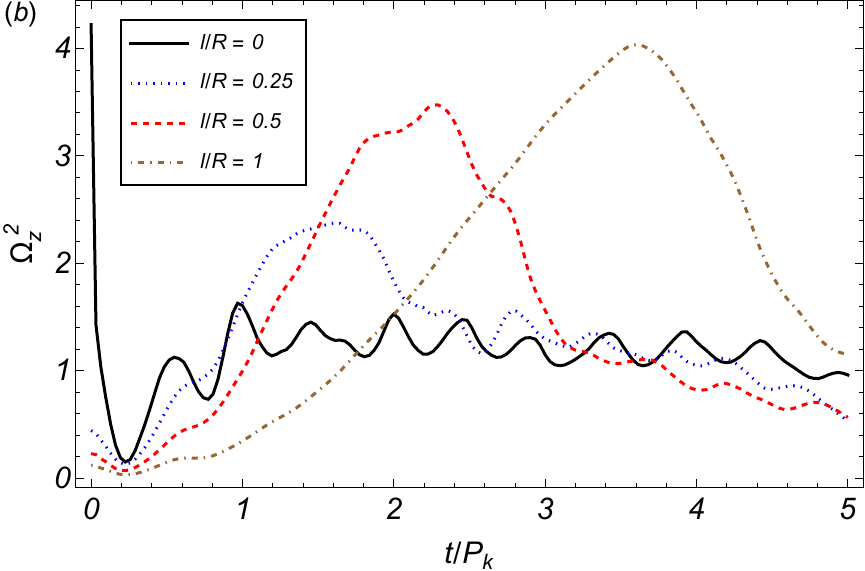}
    \caption{Weakly nonlinear simulations for the $m=2$ mode: (a) perpendicular energy integrated over the whole computational box, $E_\perp$, normalized with respect to the initial value, $E_{\perp,0}$, and (b) $z$ component of vorticity squared (in arbitrary units) integrated over the whole computational box as functions of $t/P_k$. The different line styles are for different values of $l/R$ indicated within the figure. We used $v_0/\vai = 0.05$ and $\rhoi/\rhoe = 5$ in all simulations.}
    \label{fig:ekin}
\end{figure*}

In the time-distance diagrams of Fig.~\ref{fig:timedistancem2varl}, a collective oscillation should be seen as alternating  orange and blue horizontal stripes. This is the behavior observed for the case with $l/R=0$ over practically the entire evolution, which confirms that the collective motion is dominant in this case, even if a very thin nonuniform layer is dynamically generated. The effect of resonant damping of the collective motion is seen in the case with $l/R=0.25$. Due to resonant damping, the alternating horizontal stripes  fade in time and the velocity  becomes more and more localized within the nonuniform part of the  tube. For $l/R=0.25$, the triggering of the KHi during the oscillations does not appreciably affect  the  underdamped global oscillation  predicted by the quasi-mode analysis. On the other hand, the time-distance diagram for $l/R=1$  displays vertical elongations of the horizontal stripes in the nonuniform boundary, which happen almost from the beginning of the simulation. These vertical elongations are caused by the loss of coherence of the collective motions owing to phase-mixing. Their presence supports the idea that the collective motion is so short-lived  due to the strong resonant damping when $l/R=1$ that no significant imprint of such collective behavior is left in the time-distance diagram. Finally, as was expected, the case with $l/R = 0.5$ corresponds to an intermediate situation between those of $l/R=0.25$ and $l/R=1$. The  time-distance diagram for $l/R = 0.5$ still contains a signature of the collective motion in the form of horizontal stripes at the beginning of the simulation, but  the vertical elongations of the stripes already appear after less than two periods of the global oscillation. 

In this weakly nonlinear regime, the   development of the KHi does not  appreciably affect the damping of the global oscillations, which is governed by resonant absorption. The exception is the case with $l/R = 0$, in which the resonant damping might not be fully resolved by the limited spatial resolution and the KHi probably plays a more predominant role in extracting energy from the global oscillation.

In Sect.~\ref{sec:prevtheo}, we also discussed that the density contrast, $\rhoi/\rhoe$, may play a role regarding the lifetime of the collective fluting oscillations. To study the effect of this parameter, we performed additional weakly nonlinear simulations for the $m=2$ mode in which we varied the density contrast. We considered $\rhoi/\rhoe=$~2, 5, and 10, with $l/R = 0.5$ in all cases. The corresponding time-distance diagrams  of the $x$ component of velocity along the same path as before but for these simulations are displayed in Fig.~\ref{fig:timedistancem2varrho}. The lifetime of the transient collective oscillation is very similar in the simulations with $\rhoi/\rhoe=$~5 and 10. However, a   longer collective phase is obtained in the case with $\rhoi/\rhoe=$~2. This confirms the result anticipated by Eq.~(\ref{eq:taup}) that the resonant damping timescale of the collective motion increases when  $\rhoi/\rhoe \to 1$. Nevertheless, the results for larger contrasts point out that, indeed, we must be close to that lower limit of $\rhoi/\rhoe$ for the quasi-mode lifetime to be appreciably increased.

\subsection{Exploring the triggering of the KHi}

Here, we turn to analyzing the onset of the KHi, which is a predominant effect observed during the fluting oscillations.  We note that the considered numerical resolution is not enough to  follow the full dynamics after the KHi evolves nonlinearly and turbulence sets in. Quickly, energy cascades toward small spatial scales that are below the grid resolution. A much finer resolution or an adaptive mesh refinement strategy, like that used in \citet{diaz2021}, should be adopted to  follow the dynamics deeper into the turbulent phase. That is beyond the aim of this work and so we focus on studying the initial KHi development alone.

The KHi onset in linked to the presence of azimuthal shear flows \citep[see, e.g.,][]{Browning1984,soler2010}. When $l/R=0$, the azimuthal velocity perturbation of the fluting eigenmode is discontinuous at the tube boundary \citep[see Fig.~2 of][]{soler2017}. This can naturally provide the necessary shear to directly drive the KHi if the velocity amplitude is large enough. Conversely, in the cases with $l/R\neq 0$, the continuous boundary is less prone to developing a direct instability, although this may still happen for sufficiently thin transitions. In the case of a wide, smooth, nonuniform boundary, a direct instability is less likely and the  azimuthal flows  associated with the concurrent processes of resonant absorption and phase mixing  \citep[see][]{solerterradas2015}  would be the ones that eventually trigger the KHi. 


One property of the simulations is that the KHi appearance is delayed when $l/R$ increases. This result is in agreement with the findings of \citet{diaz2021} in the case of the  KHi  caused by phase mixing of torsional ($m=0$) Alfv\'en waves. They found that the larger $l/R$, the smaller the amplitude of the phase-mixing shear flows generated after a certain  time. As a consequence of this, the larger $l/R$, the later the KHi appears and grows. However, this pattern is not satisfied in the special case with $l/R = 0$, in which a direct instability is likely to happen. In that case, the KHi is triggered at a later time than in the simulation with $l/R = 0.25$ but earlier than in the simulation with $l/R = 0.5$,  suggesting a different inception of the instability that is not related to the phase mixing flows.

The KHi onset time can  be discussed in relation with the fluting oscillation damping time. In the cases with $l/R = 0.25$ and $l/R = 0.5$, the KHi is triggered before the global fluting oscillation has  completely decayed by resonant absorption. Conversely, in the case with $l/R=1$ the KHi onset occurs  after the global oscillation has already practically stopped. At the particular time displayed in the still image of Fig.~\ref{fig:simulm2varl}, the KHi has already been triggered in all simulations except in that with $l/R=1$. The fact that in tubes with sufficiently thick  boundaries the KHi appears after the global oscillation has already been damped reinforces the idea that the KHi triggering is actually linked to the local shear flows caused by phase mixing in the wide nonuniform boundary and not to the shear associated with the collective motion of the flux tube. Therefore,  a direct instability does not occur when the nonuniform boundary is thick enough.

A more quantitative discussion of the KHi onset can be performed by studying the temporal evolution of the perturbations energy.  A convenient quantity whose evolution is informative is the sum of the kinetic and magnetic energies computed using only the perpendicular components of velocity and magnetic field. We call this quantity the perpendicular energy, for simplicity. The integral of this perpendicular energy  in the whole computational domain is
\begin{equation}
   E_\perp= \int\left[ \frac{1}{2}\rho \left( v_x^2 + v_y^2 \right) + \frac{1}{2\mu_0} \left( B_x^2 + B_y^2 \right) \right] {\rm d}{\bf r},  
\end{equation}
where  ${\bf r}$ is the position vector. Figure~\ref{fig:ekin}(a) displays the integrated perpendicular energy as a function of time for the simulations with the various values of $l/R$. For the simulations with $l/R \neq 0$, the integrated  energy is roughly constant initially, which indicates the  conservation of the energy.  However, an important energy loss does happen from a certain time onward. This occurs when the KHi is triggered and the associated turbulence sets in, so that small spatial scales below the grid resolution are quickly generated. The larger $l/R$ is, the later this phenomenon happens.  Although the loss of  energy is a purely numerical artifact due to the limited resolution, it allows us to detect the appearance of turbulence and, indirectly, quantify the onset time of the KHi. We note that a loss of  energy may also happen when the phase mixing length scale approaches the grid resolution. However, we can discard this effect as being the main cause of the sudden loss of  energy. The phase mixing has not had enough time to develop sufficiently small spatial scales approaching the grid resolution before the KHi appears \citep[see, e.g.,][for details on the phase mixing length scale]{mann1995,solerterradas2015,diaz2021}.  We checked that the times at which the sudden energy loss occurs are consistent with the initial growth of the KHi observed in the animation of Fig.~\ref{fig:simulm2varl}. Conversely, the simulation with $l/R=0$ displays a distinct behavior characterized by a more continuous energy loss from the beginning of the simulation. When $l/R = 0$, neither resonant absorption nor phase mixing operate but the KHi already plays a predominant role from an early time.

Another useful quantity that can provide further information about the mechanism behind the KHi triggering is the vorticity, $\boldsymbol{\omega} = \nabla \times {\bf v}$, specifically its component along the background magnetic field direction, $\omega_z = \left( \nabla \times {\bf v} \right) \cdot {\bf 1}_z$. We computed the square of this quantity and integrated it in the whole computational domain; namely,
\begin{equation}
    \Omega_z^2 = \int  \omega_z^2  {\rm d}{\bf r}.
\end{equation}
Figure~\ref{fig:ekin} (panel b) displays the evolution of $ \Omega_z^2$ as a function of time. The simulations with $l/R \neq 0$ are characterized by a progressive increase in the vorticity until it saturates to a maximum value and then it decreases. The saturation time approximately coincides with the time at which the sudden energy loss discussed before happens. Therefore, it can be associated with the triggering of the KHi. The subsequent decrease in the vorticity is then caused by numerical dissipation.

\begin{figure}[!tb]
    \centering
    \includegraphics[width=0.9\columnwidth]{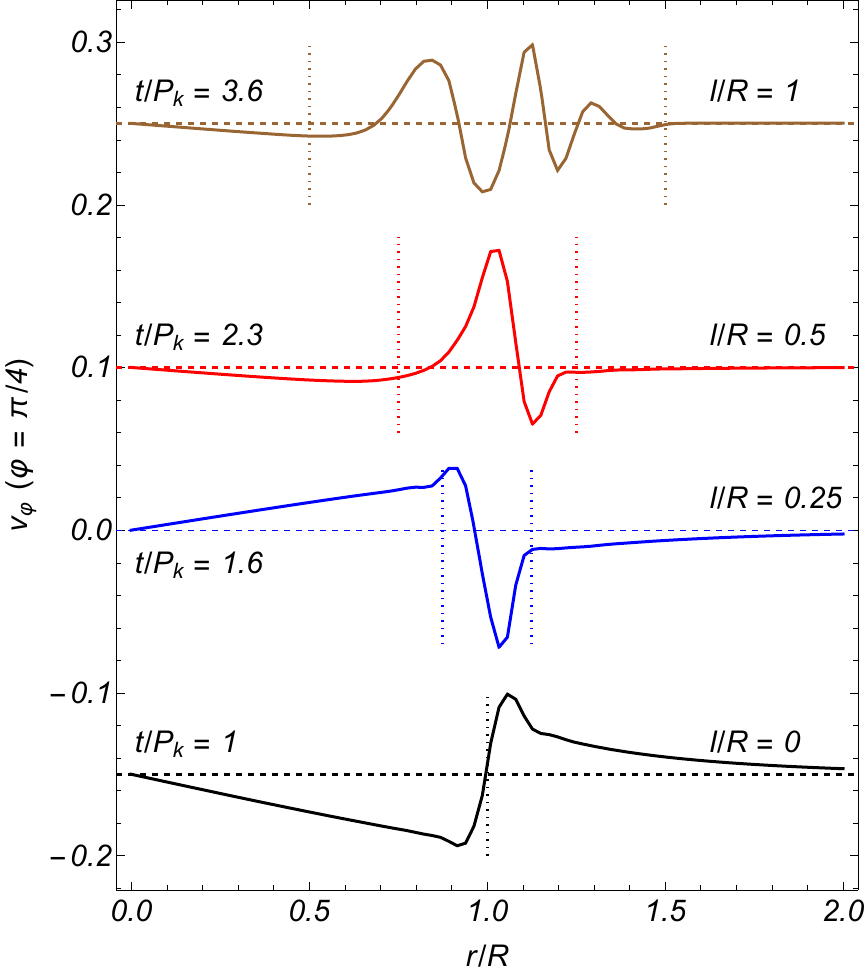}
    \caption{Azimuthal component of the velocity, $v_\varphi$, in the $z=0$ plane along a radial cut across the flux tube performed at the azimuthal angle $\varphi=\pi/4$. The azimuthal velocities for the  models with different $l/R$ have been shifted vertically by an arbitrary constant to ease the comparison. The velocities are plotted for the time  at which the maximum of the integrated vorticity occurs in each simulation, indicated within the figure. The horizonal dashed lines denote the $v_\varphi = 0$ reference value and the vertical dotted lines denote the limits of the nonuniform boundary layer in each case.}
    \label{fig:velphi}
\end{figure}


\begin{figure}[!tb]
    \centering
    \includegraphics[width=0.99\columnwidth]{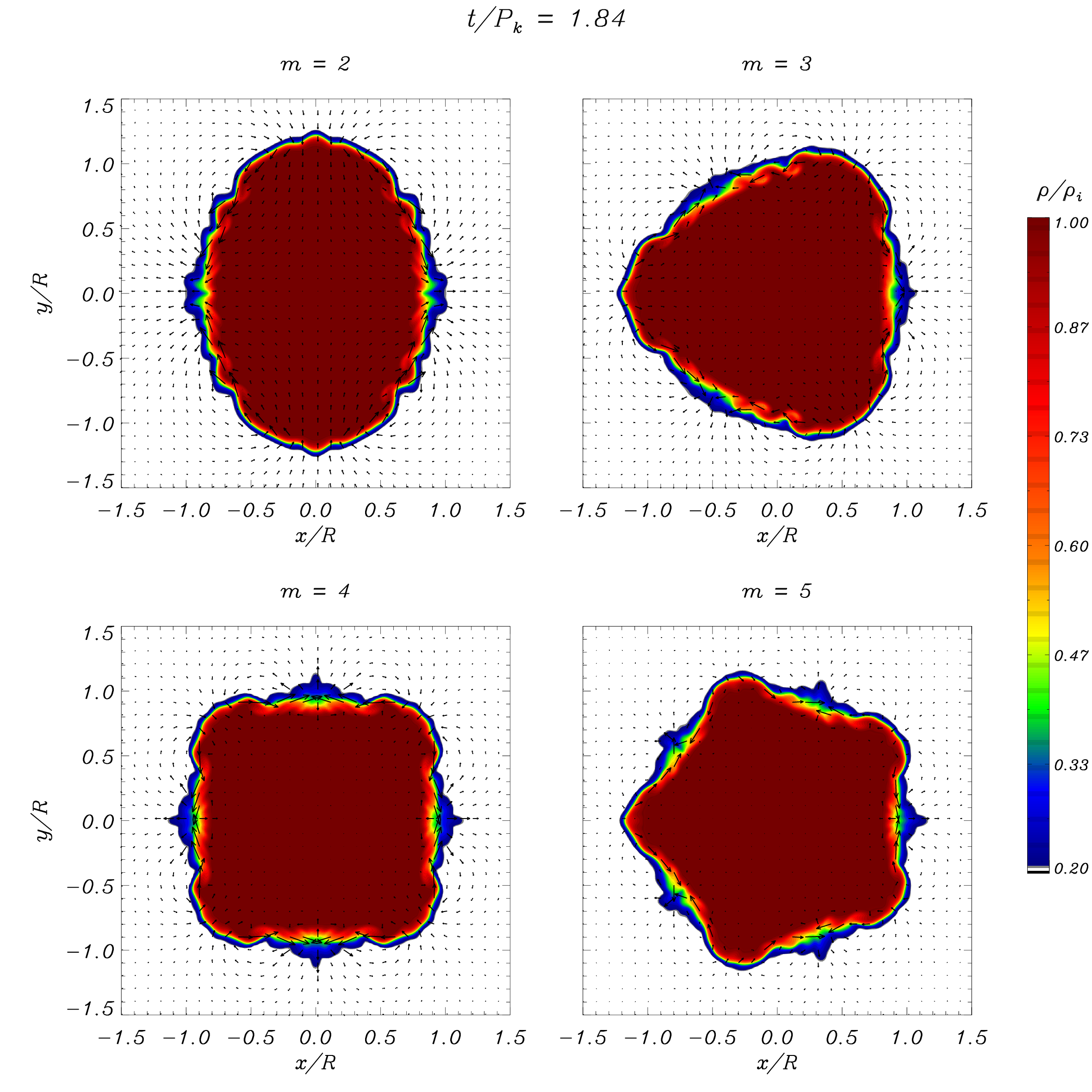} 
    \caption{Cross-sectional cuts of the density at the tube center, $z=0$, for the modes with $m=2$ (top left), $m=3$ (top right), $m=4$ (bottom left), and $m=5$ (bottom right). An initially abrupt boundary with $l/R=0$, a density contrast of $\rhoi/\rhoe = 5$, and a velocity amplitude of $v_0/\vai = 0.05$ is considered in all cases. Only a subregion of the complete numerical domain in the vicinity of the flux tube is shown. A  snapshot of the evolution at $t /P_k=1.84$ is displayed in the still image. The complete temporal evolution is available in the accompanying movie.}
    \label{fig:simulvarm0}
\end{figure}


\begin{figure}[!tb]
    \centering
    \includegraphics[width=0.99\columnwidth]{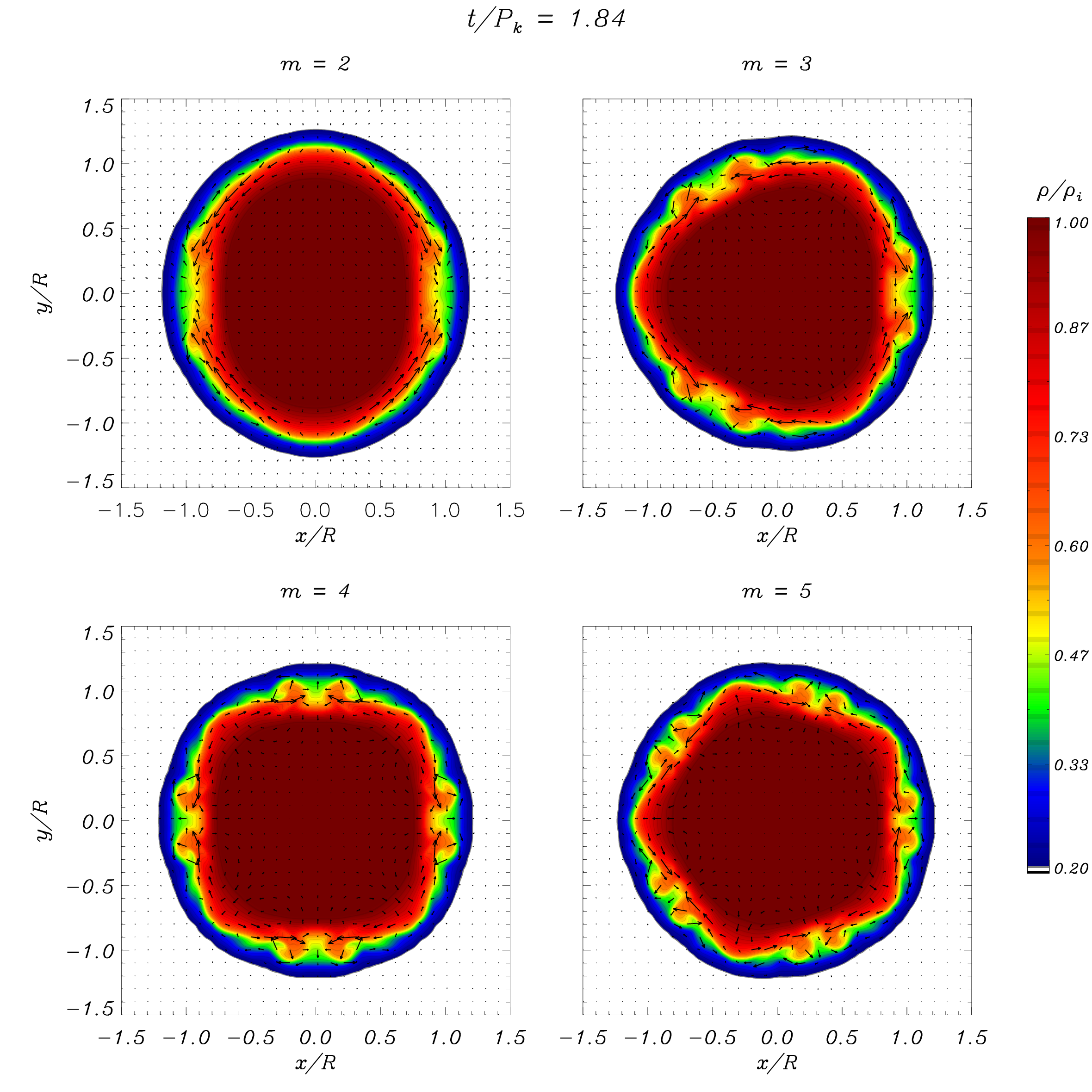} 
    \caption{Same as Fig.~\ref{fig:simulvarm0} but in tubes with an initial nonuniform boundary of $l/R=0.5$. The complete temporal evolution is available in the accompanying movie.}
    \label{fig:simulvarm}
\end{figure}

The vorticity evolution evidences how the phase mixing process progressively increases the vorticity until it is large enough to drive the KHi. An equivalent result but for torsional waves can be seen in \citet{diaz2021}. The required amount of vorticity to trigger the KHi gets larger as $l/R$ increases, signifying that it is more difficult to drive the instability in a thick boundary than in a thin one. Moreover, since phase mixing operates more and more slowly when $l/R$ increases, it takes more and more time to build up the necessary vorticity when the thickness of the  boundary increases. These two facts result in the delay of the KHi onset when $l/R$ increases.

The vorticity evolution is different in the case with $l/R=0$, owing to the different nature of the KHi driving mechanism. At $t=0$, the parallel component of vorticity is localized at the discontinuous boundary in the form of a delta function \citep[see][]{goossens2012}. When the very thin nonuniform layer is dynamically created, vorticity spreads over the nonuniform region and decreases in magnitude \citep[see][]{goossens2020}. The subsequent increase during the first period of the oscillation cannot be attributed to resonant absorption or phase mixing, since these mechanisms do not effectively work in this case, but it is more likely caused by the direct driving of the KHi. After that, the vorticity remains roughly constant.

An additional check of the KHi driving mechanism can be performed by examining the form of the azimuthal component of the velocity across the tube boundary just when the KHi is triggered.  To this end, we considered the velocity field in the $z=0$ plane, as is displayed in Fig.~\ref{fig:simulm2varl}. In a cylindrical coordinate system aligned with the axis of the flux tube,  the strongest azimuthal shear for the $m=2$ mode occurs symmetrically at the azimuthal angles $\varphi=$~$\pi/4$, $3\pi/4$, $5\pi/4$, and $7\pi/4$ defined with respect to the $x$ axis. For each one of the simulations with different $l/R$, we plotted the azimuthal component of the velocity, $v_\varphi$, along a radial cut across the flux tube performed at the angle $\varphi=\pi/4$. This was done for the simulation time  at which the maximum of the integrated vorticity occurs. These times are different for each simulation and correspond to $t/P_k \approx$~1, 1.6, 2.3, and 3.6 for the simulations with $l/R=$~0, 0.25, 0.5, and 1, respectively. These results are displayed in Fig.~\ref{fig:velphi}, where the azimuthal velocities for the different models have been shifted vertically by an arbitrary constant to ease the comparison. 

The radial profile of $v_\varphi$ when $l/R = 0$ remains close to the initial condition based on the discontinous eigenmode perturbation (Eq.~(\ref{eq:vfi1})), which has been smoothed by the dynamic generation of the thin transition. This reinforces the idea that the KHi is directly driven by the eigenmode shear when $l/R = 0$. Conversely, the radial profiles of $v_\varphi$ in the various simulations with $l/R \neq 0$ depict different stages of the phase mixing process. Remarkably, the azimuthal velocity in the $l/R = 1$ simulation displays a rather advanced stage of the process, illustrating how phase mixing builds up the necessary vorticity to drive the KHi in this thick nonuniform boundary.

\subsection{Results for different azimuthal wavenumbers}

Using the previously shown results for the $m=2$ mode as a reference, here we explore and compare the results of weakly nonlinear simulations in which we initially excite modes with other values of $m$. In addition to the $m=2$ mode, we consider $m=$~3, 4, and 5. We use $\rhoi/\rhoe = 5$ in all cases and two different values of the boundary width; namely, $l/R=0$ and $l/R = 0.5$.

Cross-sectional cuts of the density at the tube center, $z=0$, for the simulations with the four considered values of $m$ are displayed in Fig.~\ref{fig:simulvarm0} in the case with $l/R=0$ and in Fig.~\ref{fig:simulvarm} in the case with $l/R=0.5$. The transverse velocity field is also represented by means of the superimposed  arrows.  In both figures, a particular snapshot of the simulations corresponding to $t =1.84 P_k$ is shown, whereas the full  evolution can be followed in the accompanying animations.

\begin{figure}[!tb]
    \centering
    \includegraphics[width=0.9\columnwidth]{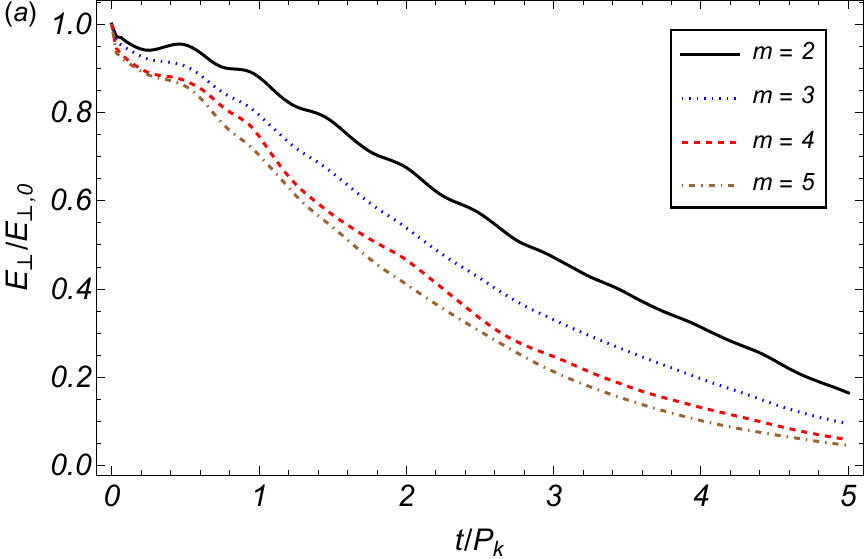} \\ \vspace{2ex}
       \includegraphics[width=0.9\columnwidth]{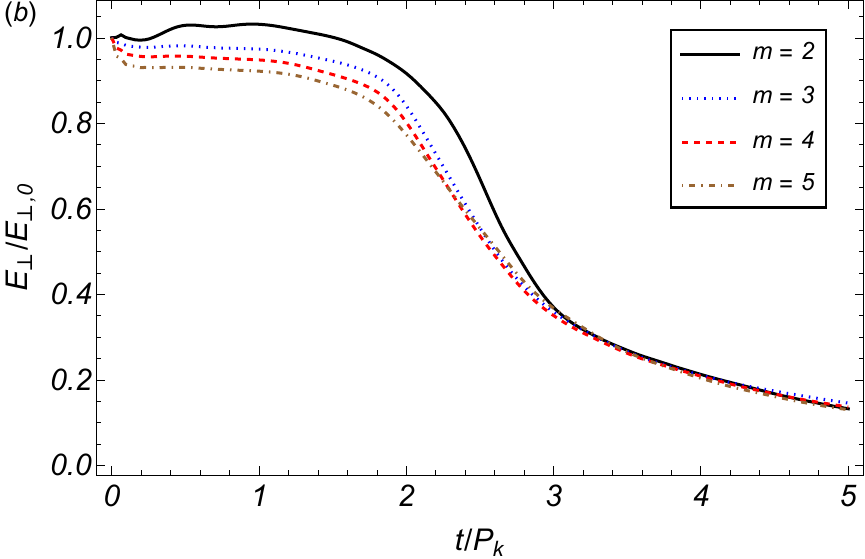} 
    \caption{Perpendicular energy integrated over the whole computational box, $E_\perp$, normalized with respect to the value at $t=0$, $E_{\perp,0}$, as a function of $t/P_k$ for the simulations with (a) $l/R=0$ and (b) $l/R = 0.5$. The different line styles are for different values of $m$ indicated within the figure. We used $v_0/\vai = 0.05$ and $\rhoi/\rhoe = 5$  in all simulations.}
    \label{fig:ekin2}
\end{figure}

The behavior of the simulations for the various values of $m$ is similar to that already discussed for the $m=2$ mode. Obviously, the most important difference resides in the azimuthal symmetry of the perturbations, but the eventual KHi onset is a common ingredient of all simulations. Visual examination of the temporal evolution when $l/R=0.5$ allows us to notice that the triggering of the KHi happens at practically the same time in the simulations with $m=$~3, 4, and 5, but it is slightly delayed in the simulation with $m=2$. The particular snapshot displayed in the still image of Fig.~\ref{fig:simulvarm} corresponds to an early time for which the KHi is clearly less developed in the simulation with $m=2$. Such a feature is not evident in the case with $l/R=0$, for which the KHi appears at approximately the same time regardless of the value of $m$ (see Fig.~\ref{fig:simulvarm0}). To evidence this finding in a more quantitative manner, Fig.~\ref{fig:ekin2} displays the perpendicular energy integrated over the whole computational domain as a function of time. When $l/R=0.5$, Fig.~\ref{fig:ekin2} shows that the loss of  energy, approximately corresponding to the KHi onset, begins at practically the same time when $m=$~3, 4, and 5, but it begins at a slightly later time in the case with $m=2$. This confirms the late onset of the KHi when $m=2$. Conversely, when $l/R=0$ the  energy decreases in a similar fashion in all cases, although it can be seen that the larger $m$ is, the stronger the energy loss is.

When $l/R=0$, the KHi is a direct instability triggered by the shear of the fluting eigenmode. As the initial velocity amplitude is the same in all cases, the shear is the same regardless of the value of $m$. This explains why the KHi onset happens at approximately the same time. On the other hand, when $l/R=0.5$ the KHi is triggered by the phase-mixing flows and to explore the physical reason why it is triggered at a later time when $m=2$, we resort again to the quasi-mode damping rate plotted in Fig.~\ref{fig:frobenius}. For $l/R=0.5$, the modes with $m=$~3, 4, and 5 have almost the same damping rate, but that for the mode with $m=2$ is larger. As a consequence of that, the resonant transfer of the global oscillation energy to the nonuniform layer is slower for the $m=2$ mode when compared to the modes with larger $m$, for which the energy transfer happens at almost the same rate. This causes the building up of vorticity owing to the phase-mixing shear flows  to happen at a  slower pace when $m=2$. In other words, it takes more time to rise the vorticity to the required levels to trigger the KHi when $m=2$ than for the other $m$'s.


\begin{figure}[!tb]
    \centering
    \includegraphics[width=0.99\columnwidth]{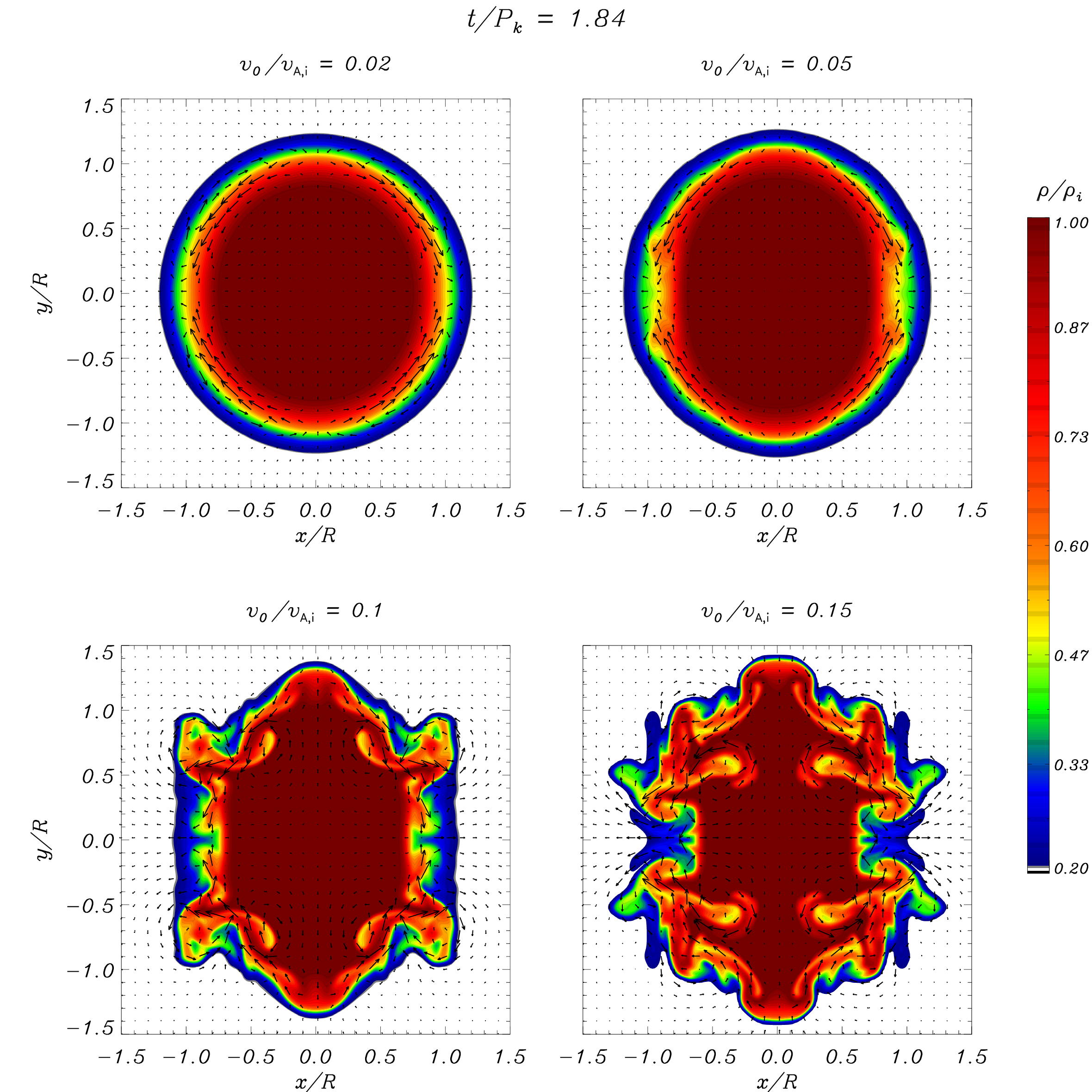} 
    \caption{Simulations of the $m=2$ mode with $l/R = 0.5$,  $\rhoi/\rhoe = 5$, and different initial amplitudes. Cross-sectional cuts of the density at the tube center, $z=0$, for the cases with $v_0/\vai = 0.02$ (top left), $v_0/\vai = 0.05$ (top right), $v_0/\vai = 0.1$ (bottom left), and $v_0/\vai = 0.15$ (bottom right). Only a subregion of the complete numerical domain in the vicinity of the flux tube is shown. A  snapshot of the evolution at $t /P_k=1.84$ is displayed in the still image. The complete temporal evolution is available in the accompanying movie.}
    \label{fig:m2nonlinear}
\end{figure}

\begin{figure*}[!tb]
    \centering
    \includegraphics[width=1.95\columnwidth]{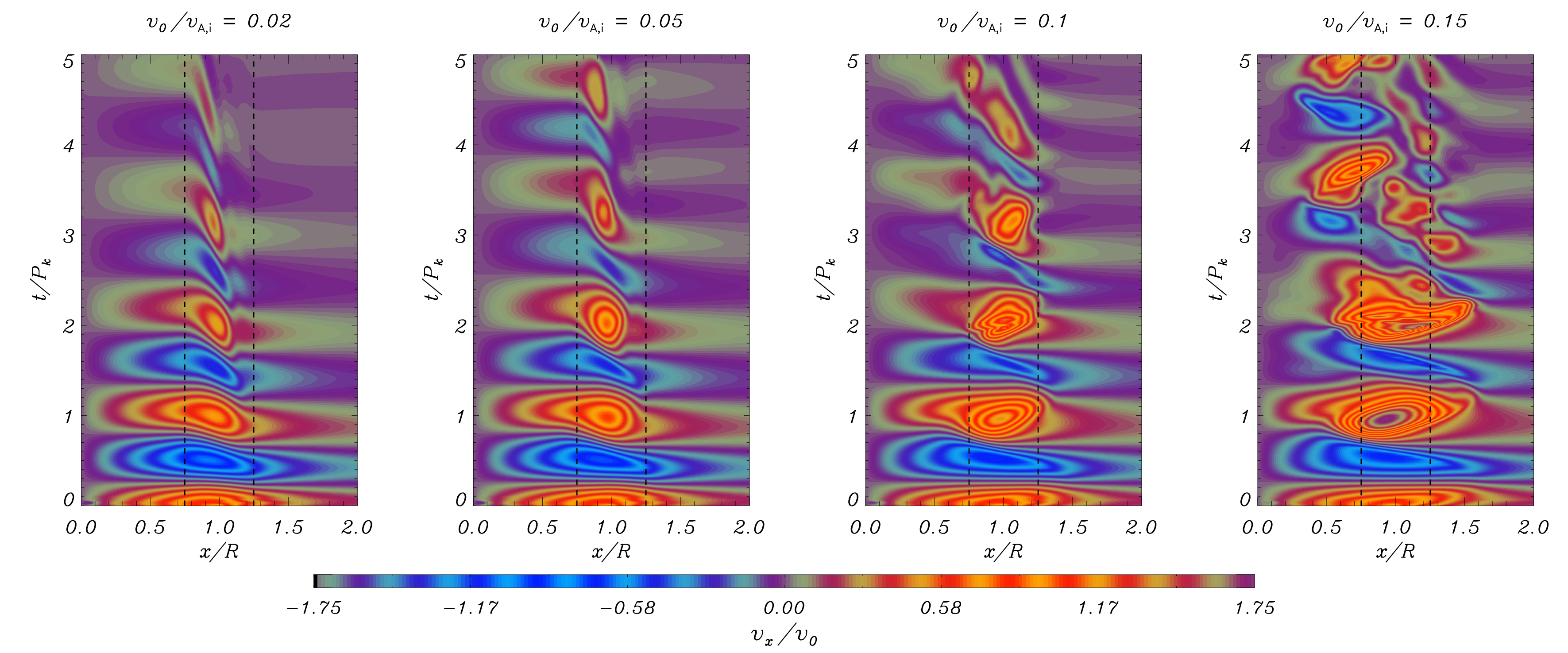} 
    \caption{Time-distance diagrams of the $x$ component of velocity at $y=z=0$ and $x\in[0,2R]$ for the simulations of the $m=2$ mode with $l/R = 0.5$ and $\rhoi/\rhoe = 5$ . From left to right, the four panels correspond to the results with $v_0/\vai =$~0.02, 0.05, 0.1, and 0.15, respectively. The vertical dashed black lines denote the boundaries of the nonuniform region in the equilibrium. The velocity is normalized with respect to the initial amplitude.}
    \label{fig:timedistancem2varv}
\end{figure*}

Another remarkable feature of the simulations discussed here is that, at certain times of the early evolution, the cross-sectional shape of the density in the inner core of the tube takes the form of quasi-regular polygons, while it remains circular in the outermost layers. This is also clearly discernible in the still images of Figs.~\ref{fig:simulvarm0} and \ref{fig:simulvarm}, where the inner core has triangular, square, and pentagonal shapes in the simulations with $m=$~3, 4, and 5, respectively. These shapes  appear briefly during the evolution and are more predominant in the case with $l/R=0$. The inner core is uniform and, therefore, phase mixing does not operate in the inner core, which  moves as a whole. In the case with $l/R=0.5$, we observe that the outermost layers of the nonuniform boundary do not follow the same shape as the inner core. This fact evidences how the different  layers in the tube become decoupled owing to transverse nonuniformity.  The result that the inner core adopts such well defined shapes, although briefly, can be attributed to the nonlinear evolution of the oscillations. The process is reminiscent to that discussed by \citet{terradas2018}  owing to the squashing of the loop produced by the inertia of the dense core in the case of kink oscillations \citep[see also][]{antolin2019}.  Nonlinearity transfers energy to higher values of $m$, which are all created in phase. This process modifies the  wave packet that initially contained only one value of $m$ imposed by the perturbation at $t=0$.  Power is concentrated at the wave nodes, which results in the steepening of the perturbations in the azimuthal direction, making the wave profiles to resemble regular shapes.  This effect can also be seen in the torsional mode simulations of \citet{diaz2021},  see their Fig.~6,  in which the initially circular velocity rings owing to phase mixing progressively acquire a squarish shape. The nonlinearity  can be quantified by comparing the amplitude of the plasma velocity  against the local Alfv\'en speed. The inner core has a larger density than the rest of the tube, which results in a lower value of the local Alfv\'en speed. Hence, for the same velocity amplitude, the oscillations tend to behave in a  more nonlinear fashion in the inner core than in the outer part where the local Alfv\'en speed is larger.

\section{Strongly nonlinear oscillations}

The amplitude of the initial velocity perturbation can have a relevant effect on the subsequent dynamics of the oscillations, since the larger the initial amplitude, the quicker nonlinear effects  become important. We considered $v_0/\vai = 0.05$ in the results shown so far. Now, we explore the effects induced by considering larger amplitudes.

\subsection{Effect of the initial velocity amplitude}

First, we focus on the $m=2$ mode in a flux tube with $l/R = 0.5$ and $\rhoi/\rhoe = 5$. We performed a series of simulations varying the initial amplitude, considering the values $v_0/\vai =$~0.02, 0.05, 0.1, and 0.15. For these simulations, the evolution of the density and the velocity field in the $z=0$ plane can be seen in Fig.~\ref{fig:m2nonlinear} and its accompanying animation. In turn, to better visualize the collective oscillation and its decay, Fig.~\ref{fig:timedistancem2varv} shows time-distance diagrams of the $x$ component of velocity along $y=0$ and for $x\in[0,2R]$.

 When $v_0/\vai = 0.02$, the resonant damping timescale  of the collective fluting oscillation remains the same as when $v_0/\vai = 0.05$. This is so because resonant absorption is, in essence, a linear process. Compared with the case with $v_0/\vai = 0.05$, the main consequence of using a lower initial velocity amplitude is  that the triggering of the KHi is delayed to a later time for which the collective motion has already decayed. So, for sufficiently small initial amplitudes, the triggering of the KHi does not happen during but after the global fluting oscillation, effectively discarding the possibility of a direct instability in this model with a relatively wide nonuniform transition. In the scenario of low amplitudes, the KHi excitation is caused by the phase mixing flows.

The use of velocity amplitudes larger than the reference value of $v_0/\vai = 0.05$ has more remarkable effects. In the cases with $v_0/\vai = 0.1$ and $v_0/\vai = 0.15$, the fluting oscillation behaves   nonlinearly almost from the beginning of the simulation and strong deformations of the density cross section are produced in the early stages of the evolution owing the growth of instabilities. A remarkable feature of the strongly nonlinear oscillations is that, contrary to the simulations with lower  amplitudes, now the induced instabilities cannot exclusively be linked with the KHi. In addition to the formation of vortices and rolls that we can clearly associate with the shear-driven KHi, other structures reminiscent of arrow heads are also formed  in specific locations of the tube boundary where the velocity field is normal to the boundary. We believe that the formation of these arrow heads is caused by a process similar to the Rayleigh-Taylor instability (RTi). This is further explored in Sect.~\ref{sec:rti}. Owing to the nonlinear development of these instabilities, the nonuniform boundary of the tube quickly becomes turbulent.  Turbulence  mixes the internal and external plasmas and so modifies the density distribution across the flux tube  more heavily than for smaller initial amplitudes. 

The presence of the RTi during the evolution of flux tube oscillations was already noted by \citet{antolin2018} and \cite{antolin2019} in the case of $m=1$ kink modes. \cite{antolin2019} discuss the growth of finger-like structures at the wake of the flux tube when it is transversely displaced owing to a kink oscillation. However, for the fluting oscillations studied here, the effect of the RTi on the global evolution appears to be more pronounced that for the kink oscillations. We find the growth of RTi structures at multiple locations of the tube boundary where the plasma acceleration is normal to the boundary. For a mode with azimuthal wavenumber $m$, this happens in $2m$ different locations of the boundary.

\begin{figure}[!tb]
    \centering
    \includegraphics[width=0.99\columnwidth]{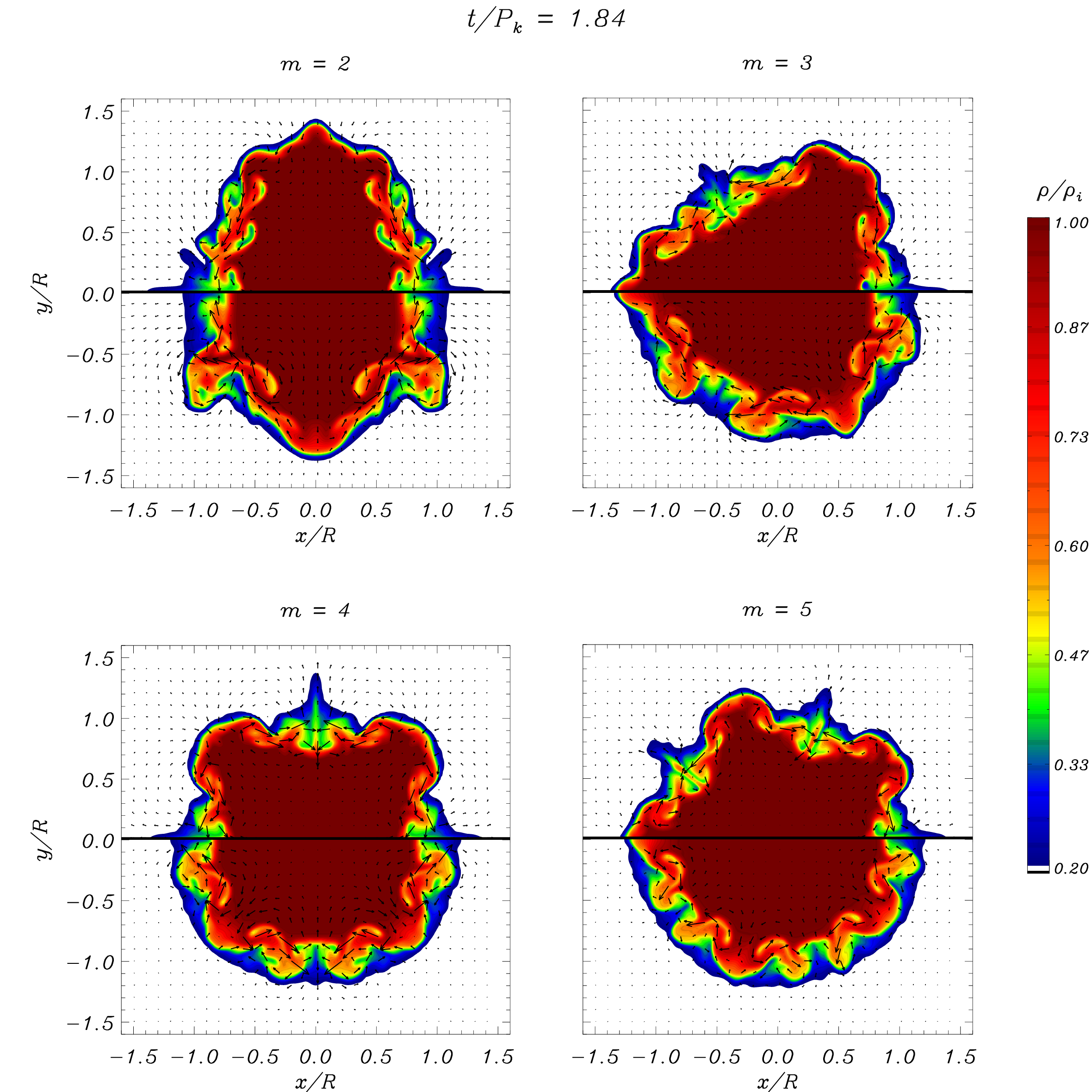} 
    \caption{
    Results of strongly nonlinear oscillations with an initial velocity amplitude of $v_0/\vai = 0.1$. Composition of cross-sectional cuts of the density at the tube center, $z=0$, for the modes with $m=2$ (top left), $m=3$ (top right), $m=4$ (bottom left), and $m=5$ (bottom right) in tubes with an initially abrupt boundary with $l/R=0$ (top half of the panels) and an     initially nonuniform boundary with $l/R=0.5$ (bottom half of the panels). A density contrast of $\rhoi/\rhoe = 5$ is considered in all cases. Only a subregion of the complete numerical domain in the vicinity of the flux tube is shown. A  snapshot of the evolution at $t /P_k=1.84$ is displayed in the still image. The complete temporal evolution is available in the accompanying movie.}
    \label{fig:simulvarm2}
\end{figure}

The growth of the RTi perturbations can be better visualized in simulations of modes with larger $m$. Figure~\ref{fig:simulvarm2} displays the results of  simulations with $v_0/\vai = 0.1$ and different azimuthal wavenumbers; namely, $m=$~2, 3, 4, and 5. In this figure, we have made a combined representation of the results obtained with $l/R=0$ and $l/R=0.5$, so that  these simulations are equivalent to those of Figs.~\ref{fig:simulvarm0} and \ref{fig:simulvarm} but with a larger amplitude. The behavior of the strongly nonlinear oscillations with different $m$'s  is qualitatively similar to that already discussed for the $m=2$ mode. In these simulations, the RTi arrow heads clearly appear at specific locations of the boundary in less than one global oscillation period, along with the formation of KHi rolls that already occurred  for lower amplitudes. The formation of the RTi structures is more predominant in the $l/R=0$ simulations. An illustration of the simultaneous formation of these two types of structures, namely KHi rolls and RTi arrow heads, in adjacent positions at the tube boundary can be seen in Fig.~\ref{fig:insta}(a), corresponding to an early phase in the simulation with $m=3$ and $l/R=0.5$.


\begin{figure}[!tb]
    \centering
    \includegraphics[width=0.85\columnwidth]{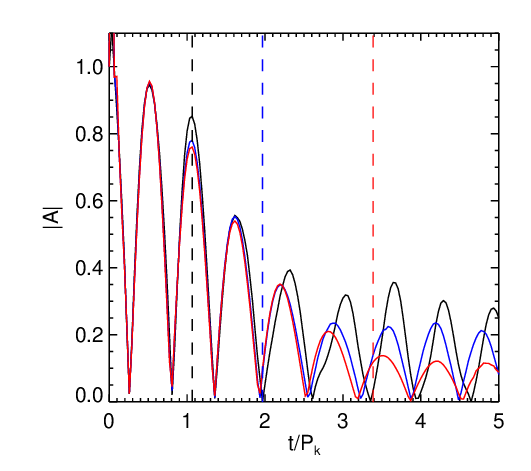} 
    \caption{$|A|$ against $t/P_k$ for simulations of the $m=2$ mode with $l/R = 0.5$ and $\rhoi/\rhoe = 5$. The red, blue and black curves show the results for $v_0/\vai =$~0.02, 0.05, and 0.1, respectively. The vertical dashed lines with the same colors show the approximate time that the KHi can be observed on the tube boundary in each case.}
    \label{fig:damping1}
\end{figure}

In the strongly turbulent scenario present when both KHi and RTi are evolving, it is difficult to quantify the efficiency of the resonant damping process. As the density profile is heavily distorted by the turbulence, the resonant energy transfer to the tube boundary may be affected. However, the turbulence itself can provide another mechanism to extract energy from the  coordinated global motion \citep[see][]{VanDoorsselaere2021}. Disentangling the resonant damping and the turbulent damping is challenging.  In a recent study on kink oscillations, \citet{hillier2023} discussed that in a heavily turbulent loop resonant absorption may be greatly reduced or even completely inhibited, while  the generation and evolution of turbulence  may entirely provide an alternative and efficient way to damp the kink oscillation. To calculate the damping for different perturbations, we looked at the dimensionless quantity, $A$, defined as
\begin{equation}
A = \frac{\partial v_x}{\partial x} \frac{R}{v_0},
\end{equation}
where $v_x$ is the $x$ component of the velocity,  and considered the results at  $r = z=0$; that is,  on the tube axis in a cross-sectional cut at the tube center. The reason for using this quantity is that, unlike kink oscillations, we cannot use the displacement of the axis. Figure \ref{fig:damping1} shows the temporal evolution of $A$ for $v_0/\vai =$~0.02, 0.05, and 0.1. For the first one and a half periods, the evolution is pretty consistent with the amplitude of the oscillation decreasing; that is, the oscillation damps. However, after this time the three simulations diverge, with the largest amplitude simulation no longer damping after about 2 periods (one period after the KHi became visible), the $v_0/\vai =$~ 0.05 simulation reaching a similar state after $\sim 2.5$ periods (approximately $3/4$ periods after the KHi became visible), and the $v_0/\vai =$~0.02 finally reaching this state after $\sim 3$ periods (again after the KHi became visible). So what is observed here is that the resonant damping of the oscillation is arrested once the resonant layer becomes turbulent.

\begin{figure}[!tb]
    \centering
    \includegraphics[width=0.85\columnwidth]{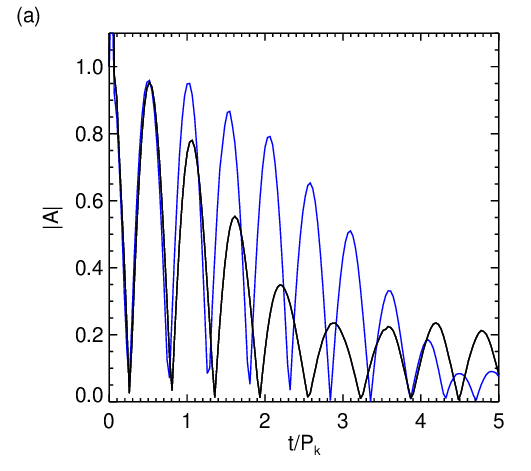} \\
    \includegraphics[width=0.85\columnwidth]{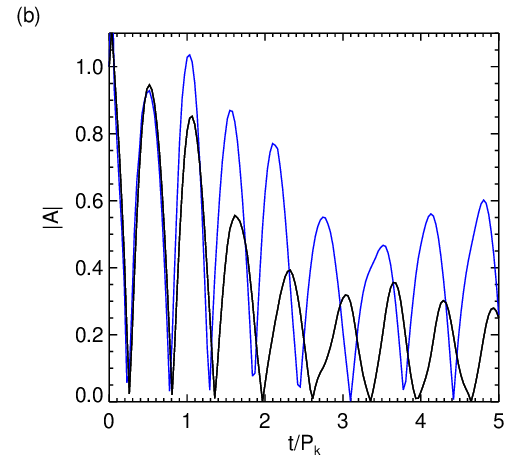} 
    \caption{$|A|$ against $t/P_k$ for the simulations of the $m=2$ mode with $v_0/\vai =$~0.05 (panel a) and 0.1 (panel b).  The blue and black curves show the results in the cases with $l/R = 0$ and $0.5$, respectively. In all cases, we used $\rhoi/\rhoe = 5$.}
    \label{fig:damping2}
\end{figure}

\begin{figure}[!tb]
    \centering
       \includegraphics[width=0.9\columnwidth]{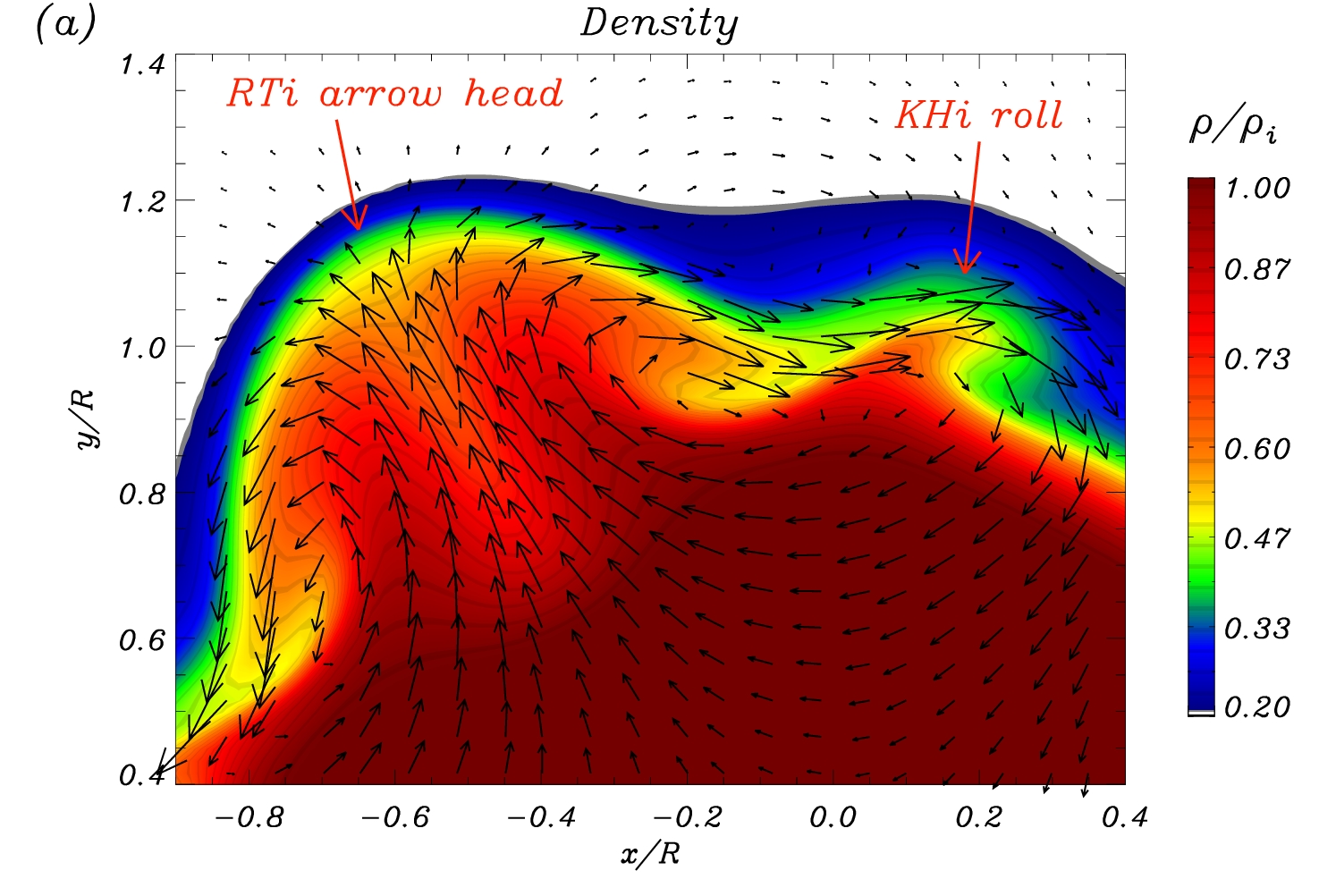}\\ 
\includegraphics[width=0.9\columnwidth]{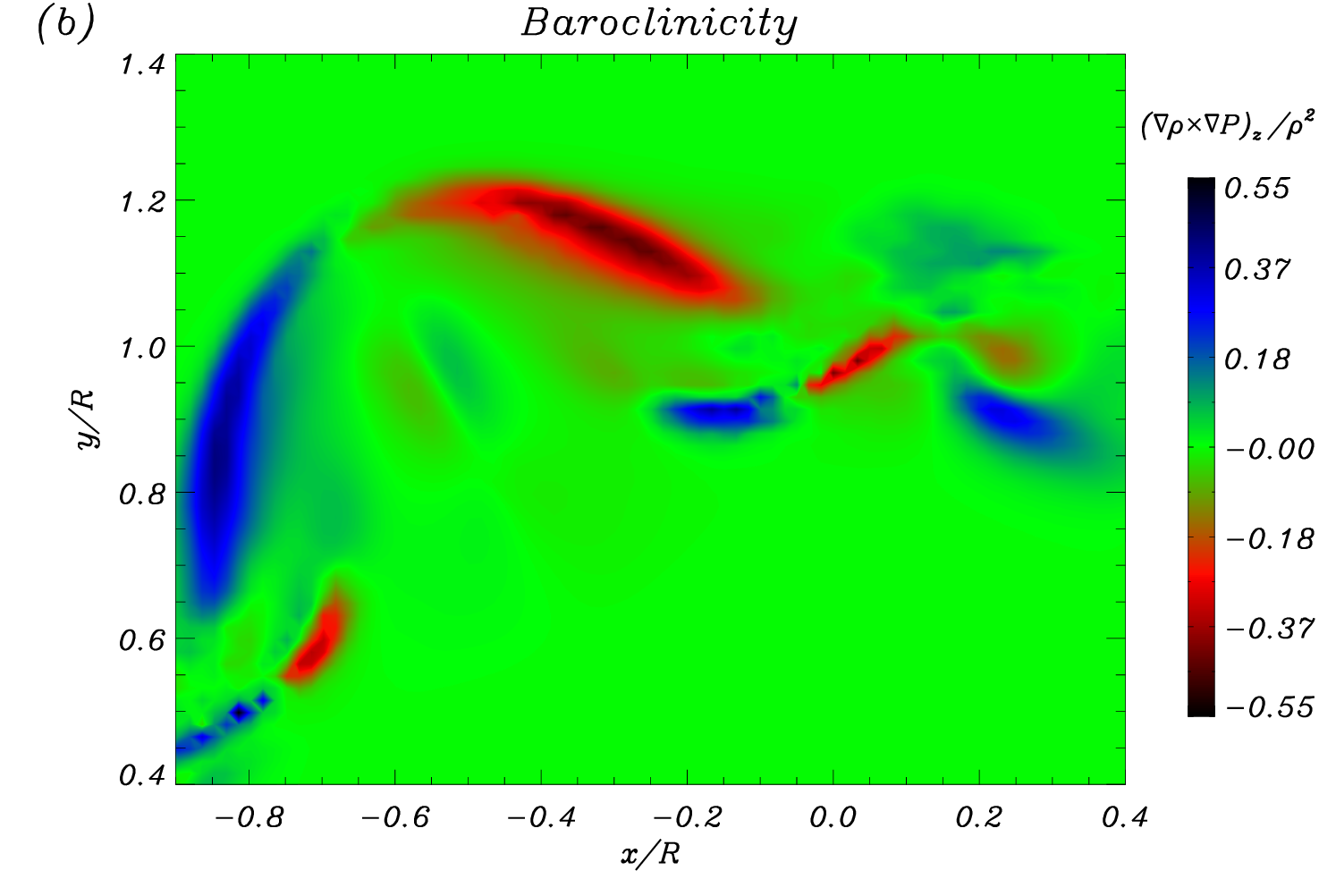}\\ 
\includegraphics[width=0.9\columnwidth]{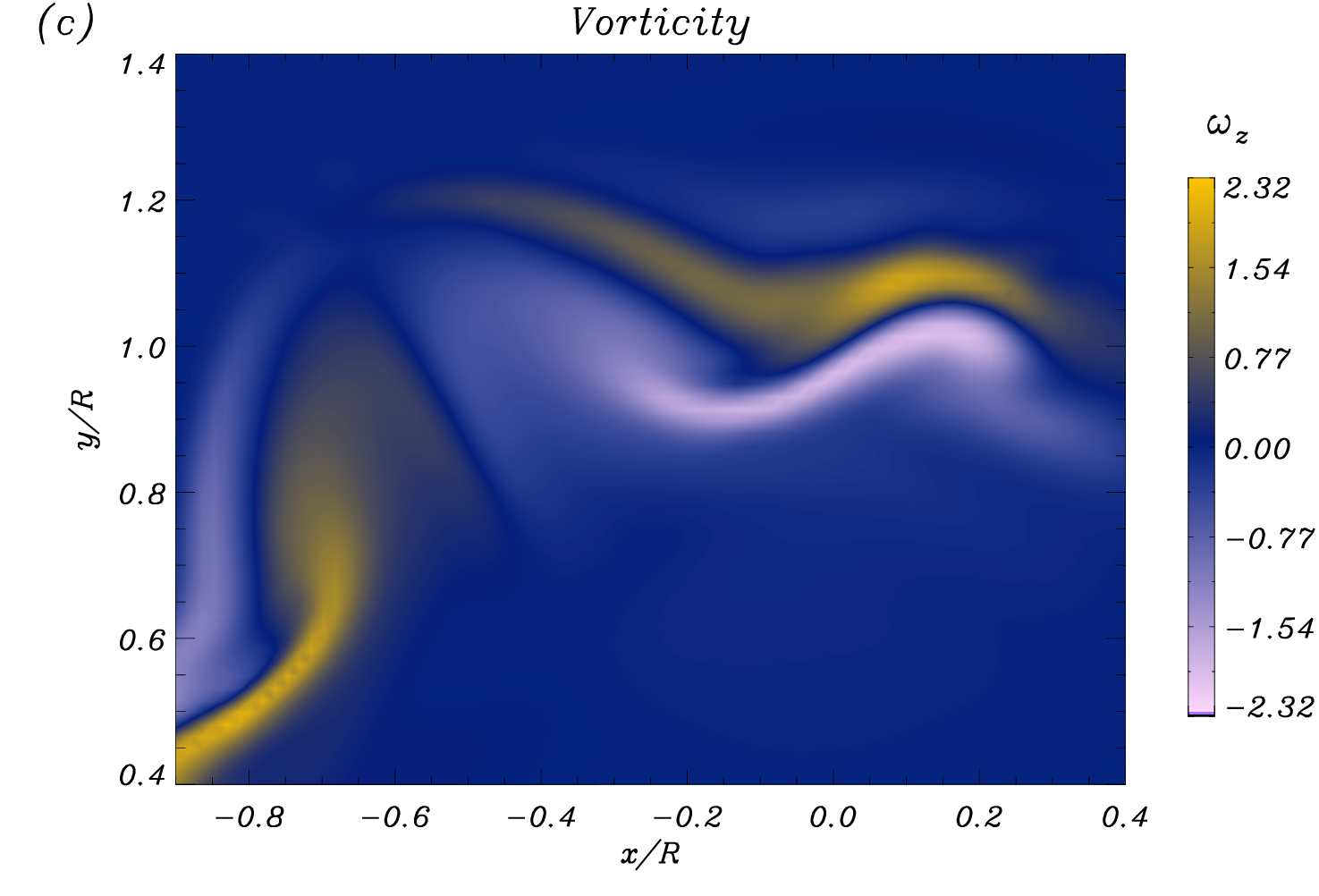}
    \caption{Close-up view of a region of the tube boundary in the strongly nonlinear simulation with $m=3$ and $l/R=0.5$, depicting the adjacent development of the KHi and the RTi: (a) density structures, (b) $z$ component of baroclinicity, and (c) $z$ component of vorticity. This simulation is the same as that shown in Fig.~\ref{fig:simulvarm2} at a time corresponding to $t/P_k = 1.1$. Baroclinicity and vorticity are given in arbitrary units. The black arrows overplotted in panel (a) denote the velocity field.}
    \label{fig:insta}
\end{figure}

If the findings of \citet{hillier2023} also apply to fluting oscillations, as the present results seem to suggest, then the damping in the heavily nonlinear regime should mostly be governed by the turbulence and not by the resonant absorption. In this line of thought, the damping of the oscillations should not depend much on the value of $l/R$ in the initial model, as Fig.~\ref{fig:simulvarm2} shows that the turbulence develops similarly in the cases with $l/R=0$ and $l/R=0.5$. According to the linear theory of resonant absorption, the simulation with $l/R=0.5$ should display a much stronger damping, but in the corresponding animation of the temporal evolution, the oscillations are seen to damp with a rather similar rate for the two cases with $l/R=0$ and $l/R=0.5$.  Using the temporal evolution of $A$ we can see the evolution of the oscillation at the core of the tube  more clearly, as is shown in Fig. \ref{fig:damping2}. Looking first at panel (a), which shows the $v_0/\vai = 0.05$ case, the simulation with $l/R = 0.5$ (black line) shows damping of the oscillation initially caused by resonant absorption, but then arrested. On the contrary, the $l/R = 0$ simulation initially shows barely no damping of the oscillation at the core of the tube and only later does the core react to the development of the KHi.  Remarkably,  at the end of the simulations, the $l/R = 0$ case has the smallest amplitude of oscillation of the two. We note that the damping envelope of the $l/R = 0$ case has some similarity to the profiles found for kink wave damping in \citet{hillier2023} and could be a signature that it can be modeled as a forced oscillator or auto-parametric resonance. More work in this direction is required. For the larger amplitude case with $v_0/\vai = 0.1$,  shown in panel (b),  the damping in the $l/R = 0$ case is still mainly governed by the KHi growth, with additional influence from the RTi. Conversely, in the $l/R = 0.5$ case the resonant damping works only initially, until it is soon disrupted by the instabilities. Then, the amplitude evolution becomes dominated by the instabilities growth and the turbulence. So, if one omits the first period in the right panel and scales the curves appropriately, it can  be seen than  the rates at which the amplitudes of the two oscillations damp at the tube core are somehow closer in evolution. This is only a qualitative comparison. Although the two simulations can have similar KHi activity, it is clear that their overall  damping profiles are different but not so different as one should expect from the linear theory of resonant damping.

\subsection{Appearance of Rayleigh-Taylor instabilities}
\label{sec:rti}

Here, we explore the appearance of the RTi. To confirm that the formation of the arrow heads is indeed caused by the development of the RTi, we can examine how the vorticity associated with the arrow head development is driven. The vorticity equation in ideal MHD is
\begin{eqnarray}
    \frac{\partial \boldsymbol{\omega}}{\partial t} + \left( {\bf v} \cdot \nabla \right)\boldsymbol{\omega} &=& \left(  \boldsymbol{\omega} \cdot \nabla \right){\bf v} - \left(\nabla\cdot{\bf v}\right)\boldsymbol{\omega} + \frac{1}{\rho^2}\nabla \rho \times \nabla P \nonumber \\
    && + \nabla \times \left[ \frac{1}{\rho} \left(  {\bf B} \cdot \nabla \right){\bf B} \right],
\end{eqnarray}
where $P$ is the total (gas plus magnetic) pressure. The third term on the right-hand side is the so-called baroclinicity, which drives vorticity because of the misalignment of  density and total pressure gradients; namely,
\begin{equation}
    {\bf b} = \frac{1}{\rho^2}\nabla \rho \times \nabla P.
\end{equation}
The classic example of the RTi development is when a heavier fluid is put on top of a lighter fluid in the presence of gravity \citep[see][]{chandra1961}, so that there is a constant acceleration (gravity) that is normal to the interface between the two fluids. The RTi that grows  in such an interface is a baroclinic instability \citep[see the review by][]{Zhou2021}. After some perturbation initially imposes a baroclinic torque, the created vorticity  will tend to further increase the misalignment of density and total pressure gradients, thus producing additional baroclinic vorticity, and so on. The RTi that appears in the fluting oscillations described here is different from the classic RTi set-up. The acceleration that is normal to the interface (the boundary of the tube) is not constant but periodic in time. Indeed, the acceleration is caused by the oscillatory nature of the flux tube motions. The development of the RTi  with a time-dependent acceleration and/or acceleration reversal  is  more complex \citep[see, e.g.,][]{dimonte2007,ramaprabhu2013,ramaprabhu2016,aslangil2016,ruderman2018,livescu2021}. However, the basic mechanism of  vorticity generation driven by the baroclinicity should be similar.

To study whether the formation of the arrow heads is actually induced by the  baroclinicity, as it should happen in a RTi, we computed the $z$ components of baroclinicity and vorticity in the same region as that displayed in  Fig.~\ref{fig:insta}(a) for the simulation with $m=3$ and $l/R = 0.5$. This is displayed in Figs.~\ref{fig:insta}(b) and (c), respectively. Large values of vorticity are generated at the locations of both the arrow head and the roll. In the case of the arrow head, we find large values of the baroclinicity of opposite signs at its sides, which confirms that the  arrow head growth is driven by the baroclinic torque. Conversely, at the position of the KHi roll the baroclinicity does not display such a strong signal as it does around the arrow head. The reason for this is that the vorticity generated at the KHi roll is driven by the shear flows and not by the baroclinicity.

\begin{figure}[!tb]
    \centering
    \includegraphics[width=0.9\columnwidth]{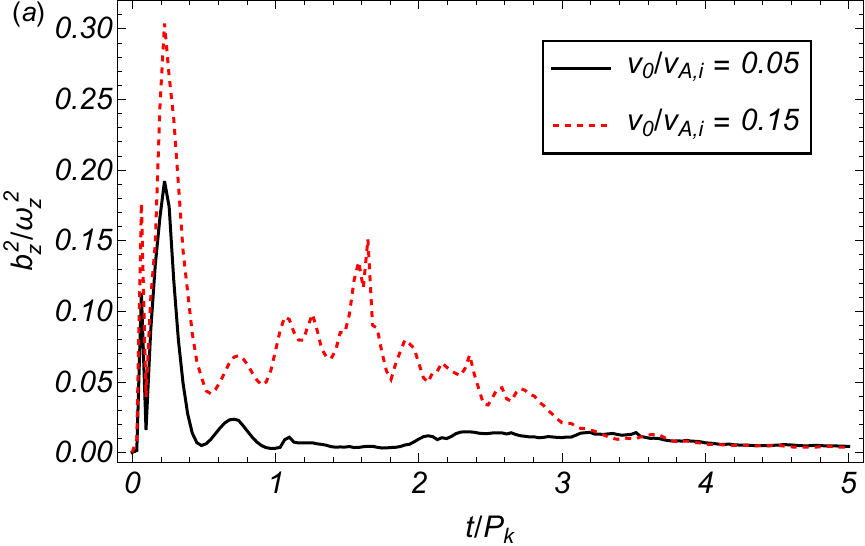}\\
     \includegraphics[width=0.9\columnwidth]{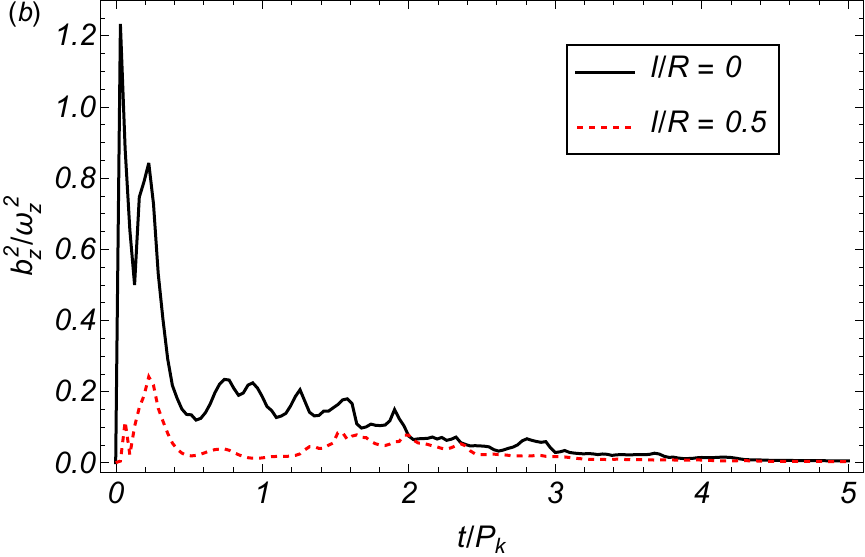}
    \caption{Ratio in arbitrary units of the $z$ components of baroclinicity squared, $b_z^2 = \left| \left( \nabla\rho \times \nabla P \right)_z / \rho^2 \right|^2$, and vorticity squared, $\omega_z^2 = \left| \nabla \times {\bf v} \right|_z^2$, integrated in the $z=0$ plane  as a function of time in simulations with  $m=2$ and: (a) $l/R=0.5$ and two different initial amplitudes, namely $v_0/v_{\rm A,i} = 0.05$ and $v_0/v_{\rm A,i} = 0.15$; (b) $v_0/v_{\rm A,i} = 0.1$ and two different thicknesses of the nonuniform layer, namely $l/R = 0$ and $l/R=0.5$.}
    \label{fig:baro}
\end{figure}

Necessarily, the temporal scale associated with the oscillating acceleration (in this case, the period of the flux tube oscillation) needs to be longer than the linear growth time of the RTi for the instability to be able to develop. In our configuration, the amplitude of the flux tube oscillations  plays an important role in setting the amount of energy that is initially deposited in the RT unstable modes, and the RTi growth time must depend on various parameters, which may include the density contrast and the thickness of the nonuniform layer. If the initial amplitude is too small, the RTi perturbations would not have enough time to  grow into the nonlinear regime before the acceleration reversal occurs,  even if the condition that the period of the oscillating acceleration is longer than the RTi linear growth time is fulfilled. This explains why the RTi is not seen in the simulations with lower amplitudes discussed in Sect.~\ref{sec:lowamp}. If the RTi is able to grow into the nonlinear regime before the acceleration reversal, then the reversal of the acceleration direction strongly affects the nonlinear development of the RTi structures, resulting in a very different evolution compared with the classic RTi. 

A thorough investigation of the RTi driven by fluting oscillations requires a  detailed analysis beyond the aims of the present study. However, a simple illustration of the effects of the initial amplitude and the thickness of the nonuniform layer on the RTi development can be seen in Fig.~\ref{fig:baro}. To estimate the relative importance of the baroclinicity in the generation of vorticity, we computed the ratio of the $z$ components of baroclinicity squared and vorticity squared, and integrated the result in the $z=0$ plane. We call this quantity the baroclinic ratio, for simplicity. The output of such a calculation is plotted against time for the $m=2$ mode for  different values of the initial amplitude and thickness of the nonuniform layer. In Fig.~\ref{fig:baro}(a) the baroclinic ratio is compared for two  values of the initial amplitude, $v_0/v_{\rm A,i} = 0.05$ and $v_0/v_{\rm A,i} = 0.15$, while $l/R = 0.5$ in both cases. The baroclinic torque  produced  by the oscillating flux tube in the initial stages of the dynamics is seen as an initial increase in the baroclinic ratio. Then, the baroclinic ratio decreases to negligible levels in the simulation with $v_0/v_{\rm A,i} = 0.05$, where the RTi is not seen to grow. Conversely, the baroclinic ratio remains significant in the simulation with $v_0/v_{\rm A,i} = 0.15$ during the first 3 periods of the fluting oscillation. It is during that time span that the RTi is seen to develop in the simulation. On the other hand, Fig.~\ref{fig:baro}(b) shows the baroclinic ratio for $l/R = 0$ and $l/R=0.5$, while $v_0/v_{\rm A,i} = 0.1$ in the two cases. Although the RTi is seen to develop in both simulations, it has a more pronounced effect on the evolution in the case with  $l/R = 0$. This translates to a larger baroclinic ratio when $l/R = 0$ than when $l/R=0.5$. The fundamental reason for this diference resides in the fact that the RTi is able to grow faster at the abrupt interface of the $l/R = 0$ case than at the smooth transition of the $l/R=0.5$ case.

\section{Conclusion}

The main purpose of this paper has been to describe the temporal evolution of Alfv\'enic fluting oscillations in transversely nonuniform magnetic flux tubes using  MHD numerical simulations. Previous works focused on studying kink and torsional modes, but a dedicated investigation into fluting modes was absent.

First, we have confirmed the earlier results by \citet{soler2017} in linear MHD about the overdamped nature of the fluting oscillations in tubes with smooth nonuniform boundaries. Underdamped fluting oscillations can only occur in tubes with thin nonuniform transitions and quite low-density contrasts. Later, the nonlinear evolution of fluting modes has been studied, which displays the same essential features of the kink mode evolution that has been extensively investigated in the literature \citep[see, e.g.,][among others]{terradas2008,antolin2014,howson2017,hillier2019,antolin2019}. These common ingredients  include resonant damping, phase mixing, KHi onset and growth, and turbulence generation. Some  differences with kink modes are that the flux tube axis is not displaced, the resonant damping  is faster, and the plasma mixing that follows the KHi and turbulence development has a different pattern in accordance with the azimuthal symmetry of the modes. Apart from that, the present results highlight that the complex dynamics of kink modes  is not unique to them but  shared among all types of Alfv\'enic modes  in solar atmospheric flux tubes.

The simulations shown here also reveal the coexistence of the KHi and the RTi during the  evolution of strongly nonlinear fluting oscillations. The presence of the RTi was already seen in kink mode simulations, but it appeared to have a minor influence on their evolution compared to that of the KHi \citep[see][]{antolin2018,antolin2019}. Conversely, highly nonlinear fluting oscillations are strongly affected by both KHi and RTi. The simulations show that the two instabilities appear at different locations of the tube boundary.  The KHi is driven at locations with strong azimuthal shear flows. In turn, the RTi shows up at locations where the plasma acceleration is normal to the boundary and is driven by the baroclinic torque related to the misalignment of density and total pressure gradients.  The development of these instabilities is characterized by the oscillating nature of  the shear flows and the normal acceleration.

In the case of heavily nonlinear oscillations, the joint effect of both KHi and RTi sets complicated turbulent dynamics that may impact on the efficiency of the resonant damping \citep{hillier2023}. However, it is generally found that the oscillations remain strongly attenuated in this scenario too, at least in most parts of the evolution.  In addition to arresting the resonant damping,  the turbulence can provide an alternative way to attenuate the global fluting oscillation that may even dominate over the resonant damping. This important aspect requires further study.

We have considered a moderate resolution in the simulations included here, because the aim was to perform many simulations to explore the parameter space. The considered spatial resolution, although limited, is enough to correctly describe the damping of the global modes. Conversely, the  development of turbulence is  affected by the spatial resolution, as turbulence quickly generates small scales. The detailed study of the turbulence evolution would require much higher resolutions.  In terms of the role of the numerical Reynolds number in the turbulent processes, which was first discussed by \citet{Onsager1949}, \citet{hillier2023} present a simple phenomenological model of a turbulent cascade showing that the timescale for energy to cascade to small scales can be approximated by a geometric series, which converges very quickly. This means that the dissipation timescale for turbulence in intermediate Reynolds numbers (lower resolutions)  is not significantly shorter than those at much larger Reynolds numbers (higher resolutions), even though the latter has significantly larger range of scales involved in the turbulent motions. Using higher resolution would provide more developed turbulence, but the description of the turbulent energy cascade in this study is appropriately achieved with the considered resolution. On the other hand, the KHi triggering can be delayed by a strong numerical diffusion, although we believe it is not an important effect in our simulations. Previous studies of torsional oscillations by \citet{diaz2021} (see their Fig. 12) have shown that the KHi onset time remains almost unaffected even when resolutions poorer than the one considered here are used. 

To end, we shall go back to the discussion given in the introduction regarding the observability of fluting oscillations in solar coronal loops. As a result of the combination of the  strong  damping and the induced instabilities, which heavily distort the tube boundary where fluting modes essentially live, it is unlikely that fluting modes could be observed in the corona as sufficiently enduring coherent oscillations of the flux tubes, even if future instruments have the required spatial resolution.

\begin{acknowledgement}
This publication is part of the R+D+i project PID2023-147708NB-I00, funded by MCIN/AEI/10.13039/501100011033 and by FEDER, EU. AH is supported by STFC Research Grant No. ST/V000659/1. For the purpose of open access, the authors have applied a Creative Commons Attribution (CC BY) licence to any Author Accepted Manuscript version arising.
\end{acknowledgement}

\bibpunct{(}{)}{;}{a}{}{,} 
\bibliographystyle{aa} 
\bibliography{refs} 

\end{document}